%% file: paper.tex
\documentclass[]{aa}
\usepackage[utf8]{inputenc}
\usepackage[bookmarks=false]{hyperref}
\usepackage{subcaption}
\usepackage{natbib}
\usepackage{fancyhdr}
\usepackage{aas_macros}
\usepackage{aalongtable}
\usepackage{mathtools}
\usepackage{amsmath}
\usepackage{blindtext}
\usepackage{bm}
\usepackage{hyperref}
\usepackage{xcolor}
\hypersetup{
  colorlinks   = true, 
  urlcolor     = blue, 
  linkcolor    = blue, 
  citecolor   = blue 
}
\usepackage[all]{hypcap} 

\usepackage{xcolor}

\begin{document}

\title{Spectral analysis of spatially-resolved 3C295 (sub-arcsecond resolution) with the International LOFAR Telescope}
\titlerunning{Multi-wavelength spectral analysis of 3C295 with the ILT}

\author{
Etienne Bonnassieux$^{1,2}$, 
Frits Sweijen$^{3}$, 
Marisa Brienza$^{1,2}$, 
Kamlesh Rajpurohit$^{1,2,4}$, 
Christopher John Riseley$^{1,2,5}$,  
Annalisa Bonafede$^{1,2}$, 
Neal Jackson$^{6}$,
Leah K. Morabito$^{7,8}$, 
Gianfranco Brunetti$^{2}$, 
Jeremy Harwood$^{9}$, 
Alex Kappes$^{11}$, 
Huub J. Rottgering$^3$, 
Cyril Tasse$^{10}$, 
Reinout van Weeren$^{3}$, 
}

\institute{
Universita di Bologna, Via Zamboni, 33, 40126 Bologna BO, Italy
\and
INAF, Instituto di Radioastronomia, Via Piero Gobetti, 101, 40129 Bologna BO, Italy
\and
Leiden Observatory, Leiden University, PO Box 9513, NL-2300RA Leiden, the Netherlands
\and
Th{\"u}ringer Landessternwarte (TLS), Sternwarte 5, 07778 Tautenburg, Germany
\and
CSIRO Astronomy and Space Science, PO Box 1130, Bentley, WA 6102, Australia
\and
University of Manchester, Department of Physics and Astronomy, School of Natural Sciences, University of Manchester, Manchester M13 9PL, UK
\and 
Centre for Extragalactic Astronomy, Department of Physics, Durham University, Durham DH1 3LE, UK
\and
Institute for Computational Cosmology, Department of Physics, University of Durham, South Road, Durham DH1 3LE, UK 
\and
Centre for Astrophysics Research, School of Physics, Astronomy and Mathematics, University of Hertfordshire, College Lane,Hatfield, Hertfordshire AL10 9AB, UK
\and
GEPI, Observatoire de Paris, 5 place Jules Janssen, 92190 Meudon, France
\and
Institut f{\"u}r Theoretische Physik und Astrophysik, Universit{\"a}t W{\"u}rzburg, Emil-Fischer-Str. 31, 97074 W{\"u}rzburg, Germany
}

\authorrunning{E. Bonnassieux et al.}

\input{source/Defs.tex}

\input{source/abstract.tex}

\maketitle


\input{source/intro.tex}


\input{source/sec1.tex}

\input{source/sec2.tex}

\input{source/sec3.tex}
\input{source/conclusion.tex}

\begin{acknowledgements}
The authors thank F. Massaro for fruitful discussions over the course of this work, which helped improve the form and contents of this paper. 
E. Bonnassieux, A. Bonafede, M. Brienza, C.~J.~Riseley acknowledge support from the ERC-Stg grant DRANOEL, n.714245. K. Rajpurohit and M. Brienza acknowledge financial support from the ERC Starting Grant ``MAGCOW", no. 714196. 
LKM is grateful for support from the UKRI Future Leaders Fellowship (grant MR/T042842/1). 
LOFAR (van Haarlem et al. 2013) is the Low Frequency Array designed and constructed by
ASTRON. It has observing, data processing, and data storage facilities in several countries,
which are owned by various parties (each with their own funding sources), and that are
collectively operated by the ILT foundation under a joint scientific policy. The ILT resources
have benefited from the following recent major funding sources: CNRS-INSU, Observatoire de
Paris and Université d'Orléans, France; BMBF, MIWF-NRW, MPG, Germany; Science
Foundation Ireland (SFI), Department of Business, Enterprise and Innovation (DBEI), Ireland;
NWO, The Netherlands; The Science and Technology Facilities Council, UK; Ministry of
Science and Higher Education, Poland; The Istituto Nazionale di Astrofisica (INAF), Italy.
This research made use of the Dutch national e-infrastructure with support of the SURF
Cooperative (e-infra 180169) and the LOFAR e-infra group. The Jülich LOFAR Long Term
Archive and the German LOFAR network are both coordinated and operated by the Jülich
Supercomputing Centre (JSC), and computing resources on the supercomputer JUWELS at JSC
were provided by the Gauss Centre for Supercomputing e.V. (grant CHTB00) through the John
von Neumann Institute for Computing (NIC).
This research made use of the University of Hertfordshire high-performance computing facility
and the LOFAR-UK computing facility located at the University of Hertfordshire and supported
by STFC [ST/P000096/1], and of the Italian LOFAR IT computing infrastructure supported and
operated by INAF, and by the Physics Department of Turin university (under an agreement with
Consorzio Interuniversitario per la Fisica Spaziale) at the C3S Supercomputing Centre, Italy. The VLA is operated by the US National Radio Astronomy Observatory.  The National Radio Astronomy Observatory is a facility of the National Science Foundation operated under cooperative agreement by Associated Universities, Inc.
Based on observations made with MERLIN, a National Facility operated by the University of Manchester at Jodrell Bank Observatory on behalf of STFC. We thank Anita Richards and the e-MERLIN team for
the maintenance and automatic processing associated
with the MERLIN archive. 	RJvW acknowledges support from the ERC Starting Grant ClusterWeb 804208.
\end{acknowledgements}


\bibliographystyle{aa}
\bibliography{bib}

\appendix

\section{Initial calibration model of 3C295}

\pg 
Our self-calibration processes all start from the image shown below.

\begin{figure}[h!]
	\centering
	\includegraphics[width=\linewidth]{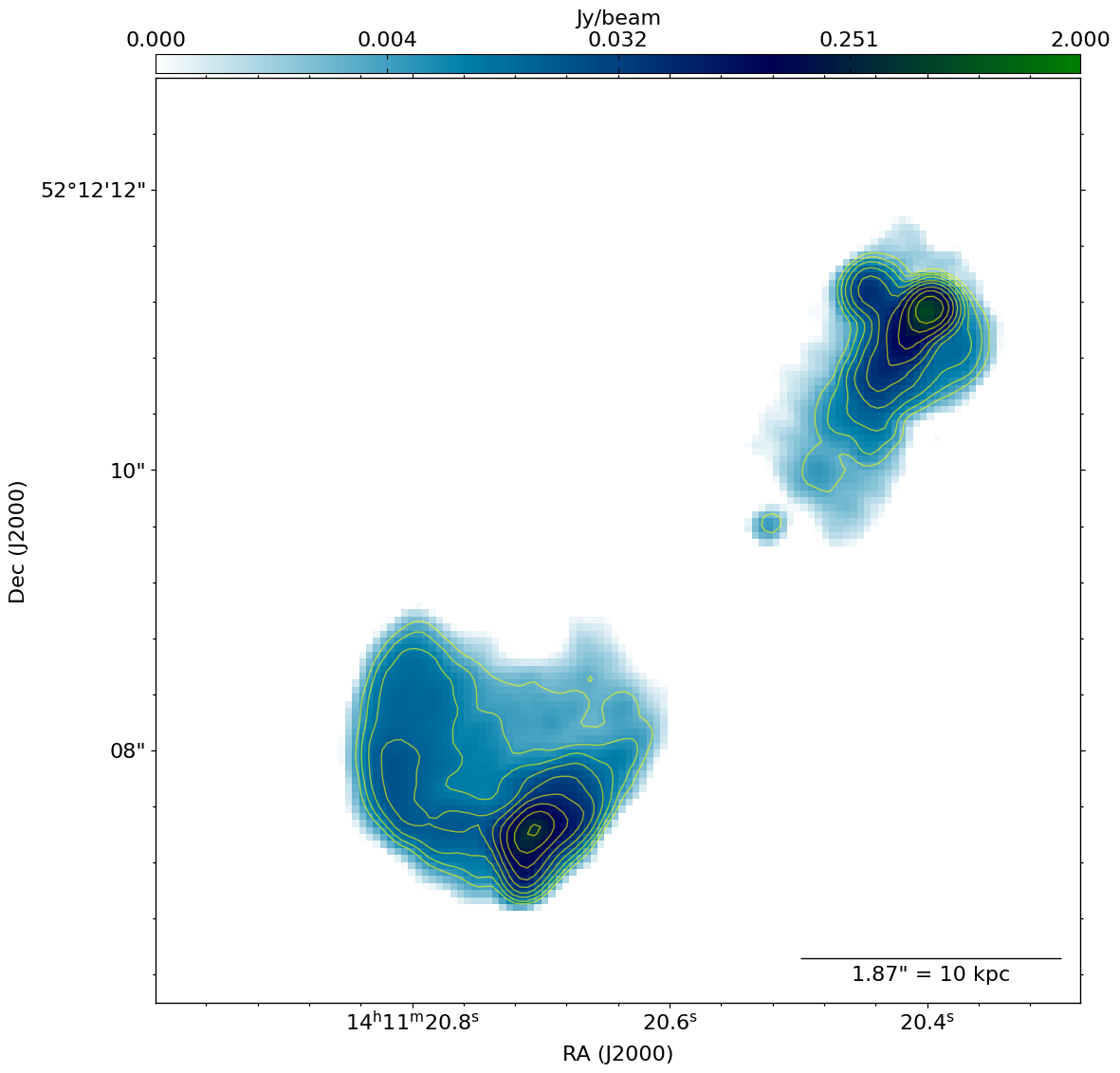} 
	\caption{VLA map of 3C295 at $8.561$ GHz, courtesy of M. Hardcastle. Pixel size is $0.05^{\prime\prime}$, beam size is $0.175^{\prime\prime}$, and flux is in Jy. The image is overlaid on itself in yellow contours starting from $0.5\sigma$ and going to 0.9 of the maximum flux present. 3C295 is clearly well-resolved and shows complex structure with both compact and diffuse emission present.} \label{fig.vla.3c295}
\end{figure}

\end{document}

%% file: source/Defs.tex
\def\pg{\paragraph{}}

\def\weight#1{\omega_{#1}}
\def\weightpr#1{\omega'_{#1}}
\def\vecweight#1{\bm{\omega}_{#1}}
\def\matJ#1{\bm{{J}}_{#1}^{t\nu}}
\def\lvec{\bm{l}}
\def\uvec#1{\bm{u}_{#1}}
\def\matK#1{\bm{{K}}_{#1,\lvec}^{t\nu}}
\def\matAa#1{{{A}}_{#1}^{t\nu}}
\def\matA#1{{{A}}_{#1,\lvec}^{t\nu}}
\def\matAo#1{{{A}}_{#1,0}^{t\nu}}
\def\matAlo#1{{{A}}_{#1,\lvec_0}^{t\nu}}
\def\k#1{k_{#1,\lvec}^{t\nu}}
\def\ko#1{k_{#1,0}^{t\nu}}
\def\klo#1{k_{#1,\lvec_0}^{t\nu}}
\def\matV#1{\bm{V}_{#1}^{t\nu}}
\def\matVhat#1{\hat{\bm{V}}_{#1}^{t\nu}}
\def\matB{\bm{B}}
\def\matBl{{\matB}_{\lvec}^{\nu}}
\def\matBlo{{\matB}_{\lvec_0}^{\nu}}
\def\I{\mathbf{I}}
\def\dirtyIpq{I_{\text{dirty}}^{pq,t\nu}}
\def\dirtyIpqsmr{\tilde{I}_{\text{dirty}}^{pq,t\nu}}
\def\dirtyI{I_{\text{dirty}}}

%% file: source/abstract.tex

\abstract{
3C295 is a bright, compact steep spectrum source with a well-studied integrated radio spectral energy distribution (SED) from 132 MHz to 15 GHz.
However, spatially resolved spectral studies have been limited due to a lack of high resolution images at low radio frequencies.
These frequencies are crucial for measuring absorption processes, and anchoring the overall spectral modelling of the radio SED. In this paper, we use International LOFAR (LOw-Frequency ARray) Telescope (ILT) observations of 3C295 to study its spatially resolved spectral properties with sub-arcsecond resolution at 132 MHz.

Combining our new 132 MHz observation with archival data at 1.6 GHz, 4.8 GHz, and 15 GHz, we are able to carry out a resolved radio spectral analysis. The spectral properties of the hotspots provides evidence for low frequency flattening. In contrast, the spectral shape across the lobes is consistent with a JP spectral ageing model.
Using the integrated spectral information for each component, we then fit low-frequency absorption models to the hotspots, finding that both free-free absorption and synchrotron self-absorption models provide a better fit to the data than a standard power law.
Although we can say there is low-frequency absorption present in the hotspots of 3C295, future observations with the Low Band Antenna of the ILT at 55 MHz may allow us to distinguish the type of absorption.


}

%% file: source/intro.tex

\section{Introduction}\label{sec.intro}

\pg
3C295 is one of the most famous active galactic nuclei in the sky, first detected in the 3C survey \citep{1959MmRAS..68...37E}. It is associated with a nearby \citep[$z=0.461$,][]{2013yCat.5139....0A} massive cD galaxy \citep{1981ApJ...251..485M}.  Its electromagnetic spectrum has been well studied from the radio \citep{1990MNRAS.244..362A,1991AJ....101.1623P, 1983IEEEP..71.1295N,2011ApJ...739L...1P}, to optical \citep[e.g.][]{1994A&A...285..785T} to X-ray \citep[e.g.][]{2000ApJ...530L..81H,2001A&A...372..755B} frequencies.
The radio integrated properties are so well known that 3C295 is a standard flux density calibrator from MHz to GHz frequencies \citep{2012MNRAS.423L..30S,2017ApJS..230....7P};
Due to its extremely high flux density (90.87 Jy at 144 MHz, 19.42 Jy at 1.5 GHz) but also compact, with a largest angular size of $6^{\prime\prime}$. Although high frequency radio observations have resolved the sub-structure of 3C295 \citep[e.g.]{1991AJ....101.1623P}, the spatially-resolved morphology at $\sim100$ MHz frequencies remains unstudied.

\pg
3C295 is a scientifically interesting source for understanding the emission and absorption processes in radio-loud AGN. 
To understand these processes, we require spatially resolved information across a broad range of radio frequencies. While optically thin radio emission can often be adequately described
using a simple power law model, sources like 3C295 can be affected by processes like spectral ageing at higher frequencies ($\gtrsim$500 MHz, depending on age) or absorption processes like free-free absorption or synchrotron
self-absorption at low frequencies ($\lesssim$200 MHz). These emission and absorption processes can be modelled, but doing so requires measurements at multiple frequencies with enough spatial resolution to
model distinct components.
While the integrated radio spectrum of 3C295 is very well-constrained and shows  a turn-over at $\sim$80 MHz \citep{2012MNRAS.423L..30S}, it is impossible to tell without spatially resolved information
at these frequencies which components of the source are driving this. 

\pg
To answer this question, we use the International LOw Frequency ARray (LOFAR) Telescope \citep[ILT, ][]{2013A&A...556A...2V} to make sub-arcsecond resolution images of 3C295 at 132 MHz for the first time.
The geographical spread of the ILT stations across Europe provide baselines up to $\sim$2,000 km, and recent advances in calibration techniques have made it easier than ever to use the array to achieve tenths of arcsecond
resolution at 150 MHz (Morabito et al., 2021). When combined with matched-resolution archival Very Large Array (VLA) and Multi-Element Radio-Linked Interferometer (MERLIN) observations, they allow us to
perform the first spatially resolved spectral modelling of 3C295 to investigate the physical conditions of its radio components.

\pg
In Sec. \ref{sec.1}, we give an overview of the data and the resulting images 
In Sec. \ref{sec.spect} we analyse the spectral properties of these images. Finally, in Sec. \ref{sec.conclusion} we close out with a discussion.
Throughout this paper, we assume a standard $\Lambda$CDM cosmological model, using \citet{2006PASP..118.1711W} to calculate distances at 3C295's redshift (Table~\ref{table.3c295.properties}). Throughout this paper, we define the spectral index $\alpha$ by $S_\nu \propto \nu^\alpha$ 
, where $S_{\nu}$ is the flux density per unit frequency $\nu$.


%% file: source/sec1.tex

\section{Observations \& Results}\label{sec.1}

\subsection{Data Reduction}

\pg
Each telescope has its own data calibration methods, and we describe our LOFAR, MERLIN and VLA data reduction separately.
Our target is one of the radio flux density calibrators described in \citet{2017ApJS..230....7P}. Although we sometimes used the flux scale described in \citet{2012MNRAS.423L..30S} during processing, which we will note when relevant in the following section, this flux scale diverges at higher frequencies, and so we align all images to the \citet{2017ApJS..230....7P} flux scale in the end. This results in an absolute flux scaling error of $5\%$ for all frequencies. 


\subsubsection{LOFAR}\label{sec.1.1}

\pg
{The LOFAR HBA data was taken during a targeted observation of 3C 295 (PI Sweijen; project code COM\_010, L693723) on January 25, 2019. It was observed from 01:41:00 to 09:40:59 UT, for a total of 8 hours on source and was book-ended by two 15 minute calibrator observations. A typical observing setup was used where the data was recorded at $3.05\ \mathrm{kHz}$ frequency resolution and $1\ \mathrm{s}$ time resolution. The data were then archived to the Long Term Archive (LTA) after they were averaged in frequency to a resolution of $12.21\ \mathrm{kHz}$ and RFI was excised using AOFlagger \cite{2010ascl.soft10017O}. The frequency coverage ranged from $120\ \mathrm{MHz}$ to $168\ \mathrm{MHz}$ for a total of $48\ \mathrm{MHz}$ of bandwidth. 51 stations in total (24 core stations (CS), 14 remote stations (RS) and 13 international stations) participated in the observation, all of which provided usable data except for the remote station RS508.}

\pg
{First the prefactor calibrator pipeline was run on one of the calibrator observations to obtain corrections for polarisation alignment, find the station bandpasses and correct for clock offsets \citep{2019A&A...622A...5D}. After this all core stations were combined into a single ``super'' station, ST001 to significantly reduce the field of view. Finally, the data were averaged to a time resolution of $4\ \mathrm{s}$ and a frequency resolution of $48.82\ \mathrm{kHz}$. The data were then self-calibrated. We used PyBDSF \citep{2015ascl.soft02007M} to generate a starting model from a VLA A-configuration image at $8.561$ GHz shown in Fig. \ref{fig.vla.3c295} \citep{2004MNRAS.351..845G}. Self-calibration then followed a strategy similar to that presented in \cite{Weeren2020}. First 5 cycles of phase-only calibration were performed at the resolution of the data. Subsequently the calibration switched to 5 cycles of phase-only calibration followed by amplitude calibration and a frequency interval of $195.28\ \mathrm{kHz}$ and a time interval of $15\ \mathrm{minutes}$. Such short intervals were possible due to the exceptionally high signal-to-noise ratio present on this class of source. Calibration was restricted to baselines $40\ \mathrm{k}\lambda$ or longer to help stabilise calibration. After each iteration a new image was made and $>7\ \sigma_\mathrm{rms}$ emission was used to update the model for the next iteration, where $\sigma_\mathrm{rms}$ is the local root-mean-square noise in the image. Imaging was done using multi-frequency synthesis, a robust $-1$ weighting and multi-scale clean. WSClean \citep{2014MNRAS.444..606O} was used for imaging and DP3 \citep{2018ascl.soft04003V} was used for calibration. Finally, we note that international LOFAR data can be impacted by large dispersive delays, which are challenging to remove for low signal to noise sources where large bandwidths are necessary to find good solutions. Fortunately, 3C295 has a high enough flux density that we can perform self-calibration per channel, removing the need to solve for these dispersive delays separately.}

\pg
This data was {further} self-calibrated using the Wirtinger pack software \citep{2015MNRAS.449.2668S,2018A&A...611A..87T}, which consists of the \texttt{DDFacet} imager and \texttt{killMS} calibration software. 

We used small intervals in time and frequency (8 seconds and 4 channels, respectively) so as to maximise our tracking of underlying changes in the antenna gains, solving for full-Jones matrices in each interval. Intervals of increased ionospheric activity (corresponding to noisier measurements and thus worse S/N for our gain solutions) were corrected for by using the quality-based weighting scheme developed in \citet{2018A&A...615A..66B}. This was key in maximising the relative contribution of the international baselines to the final image. Finally, drifts in the gain were corrected for by fixing the integrated flux density of the model to the \citet{2012MNRAS.423L..30S} flux density scale before calibration at every self-calibration pass.

\pg
The final LOFAR image was made with an inner $uv$-cut of $20k\lambda$ (shortest baseline in the data set was $\sim 950\lambda$), a Briggs weighting with robust parameter of $-2$ \citep{1995AAS...18711202B}, a $uv$-taper with FWHM of $0.6^{\prime\prime}$, and a pixel size of $0.1^{\prime\prime}$. This ensures that the effective $uv$-coverage of all observations can be matched as much as possible. This image was made using \texttt{wsclean2.10.0}'s \citep{2014MNRAS.444..606O} multi-scale CLEAN algorithm \citep{2017MNRAS.471..301O}, as DDFacet lacked an implementation of $uv$-tapering at the time of reduction. Note that all other images used exactly the same parameters.

\pg
The final image sensitivity is $0.6\rm \,mJy\,beam^{-1}$ at 132 MHz (which corresponds to a dynamic range - DR - of 13000). We note that our noise levels are nevertheless still determined by DR limitations.

\subsubsection{MERLIN}

\pg
The archival MERLIN observation was taken on 26 May 1998,
with a central observing frequency of $1.658$ GHz and a total bandwidth of $26$ MHz. The observation included 6 antennas. The data were extracted from the MERLIN archive, and flux-calibrated with 3C286. We then performed 8 passes of self-calibration using \texttt{killMS} and \texttt{DDFacet}  to improve the calibration, followed by a final re-imaging using \texttt{wsclean2.1.0}, using exactly the parameters and initial model as for our LOFAR data.

\pg
The shortest baseline available in our MERLIN observation is $25~\text{k}\lambda$, which corresponds to an angular scale of $8.25^{\prime\prime}$. It therefore does not lead to loss of flux, as 3C295 has a LAS of $6.1^{\prime\prime}$. Using the same imaging procedure as for LOFAR ensures homogeneous $uv$-coverage for the angular scales of interest. 

\pg
The final noise level achieved was $25$\,mJy\,beam$^{-1}$ , leading to a dynamic range of about 200 in the final science image. The hotspot positions were used to align the final image with the LOFAR images. This correction of $\sim 0.4^{\prime\prime}$ was done in the image plane.

\subsubsection{VLA}

\pg
Our VLA data were acquired from the NRAO Science Data Archive. We restricted our search to observations from A-configuration and longer than two hours in duration to maximise our $uv$-coverage. 
We used two observations from the NRAO Science Data Archive, each in A-configuration and longer than two hours in duration to maximise our $uv$-coverage and each comprising two frequency bands, taken as part of project AP135 (PI. R. Perley) on 12 September 1987.
The central frequencies of each observation were 4.76 GHz and 14.975 GHz, with bandwidths of 50 MHz.
Both use the full 27 VLA antennas
. This configuration leads to shortest baselines of $9.3\text{k}\lambda$ at 5 GHz and $35\text{k}\lambda$ at 15.5 GHz, corresponding to LASs of $\sim22.2^{\prime\prime}$ and $\sim3^{\prime\prime}$, respectively. Because the latter is smaller than the LAS of 3C295, we added a calibrated C-configurationd dataset from the VLA science data archive in order to properly sample 3C295 at 15GHz.

\pg
The archival data sets contained uncalibrated visibilities. We therefore used the Wirtinger pack for our self-calibration, with the exact same procedure and initial model as for LOFAR. We then imaged our final visibilities using \texttt{wsclean} to match the imaging parameters used at all other frequencies. This converged quickly for both frequencies, allowing us to acquire science images with rms noise levels of $15$~mJy beam$^{-1}$ and $670~\mu$Jy beam$^{-1}$ at 5 and 15 GHz respectively (corresponding to dynamic ranges of 165 and 754, respectively).
The hotspot positions were used to align the final image with the LOFAR images, as for the final eMERLIN image. This offset was $\sim 0.2^{\prime\prime}$.

%% file: source/sec2.tex
\subsection{Radio Images}

\begin{figure*}
	\centering
	\begin{subfigure}{.49\textwidth}
		\resizebox{\hsize}{!}{\includegraphics[width=.32\linewidth]{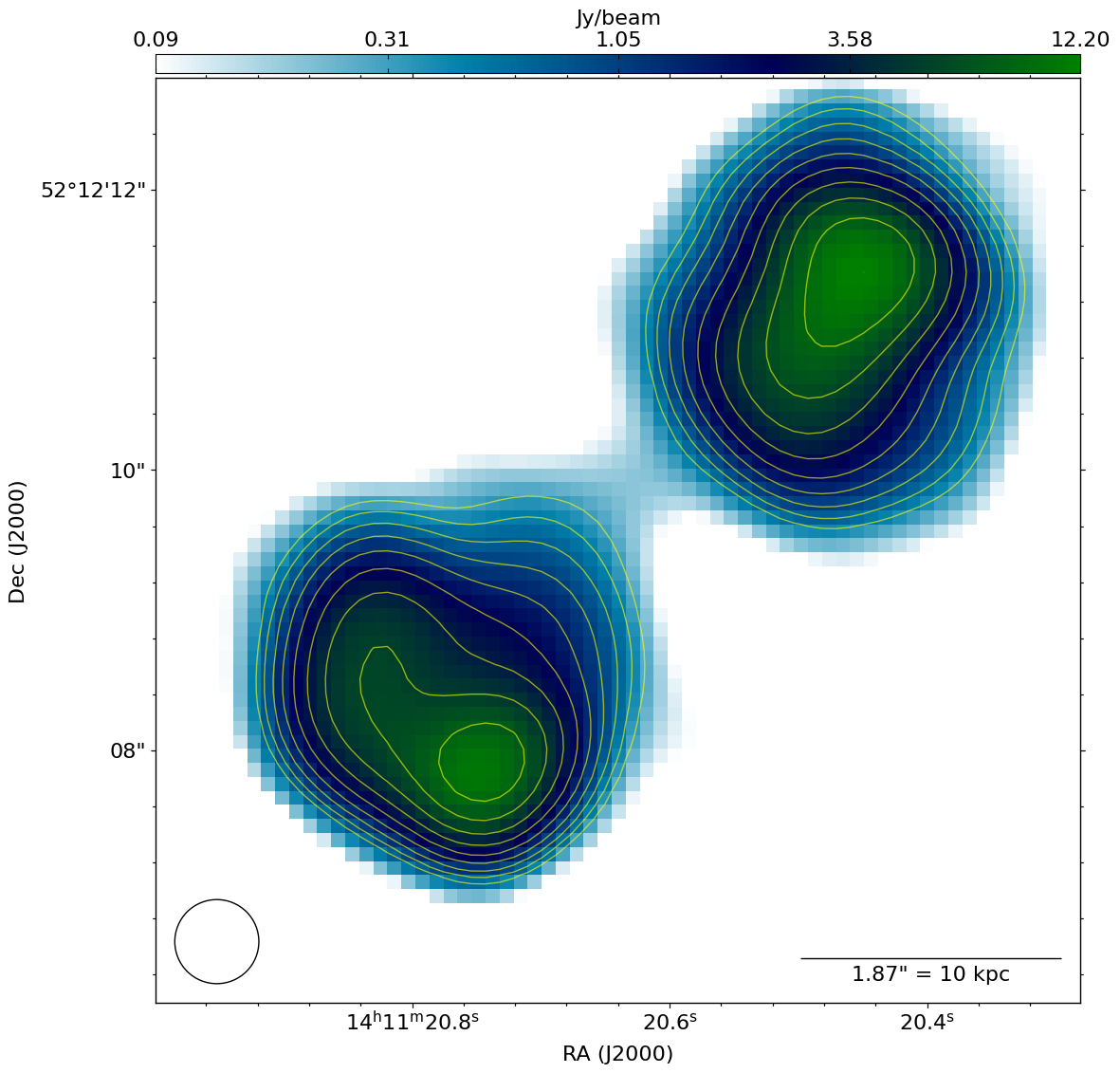}}
		\caption{LOFAR, $132$ MHz} \label{fig.imsci.lofar0}
	\end{subfigure}
 	\hfill
	\begin{subfigure}{.49\textwidth}
		\resizebox{\hsize}{!}{\includegraphics[width=.32\linewidth]{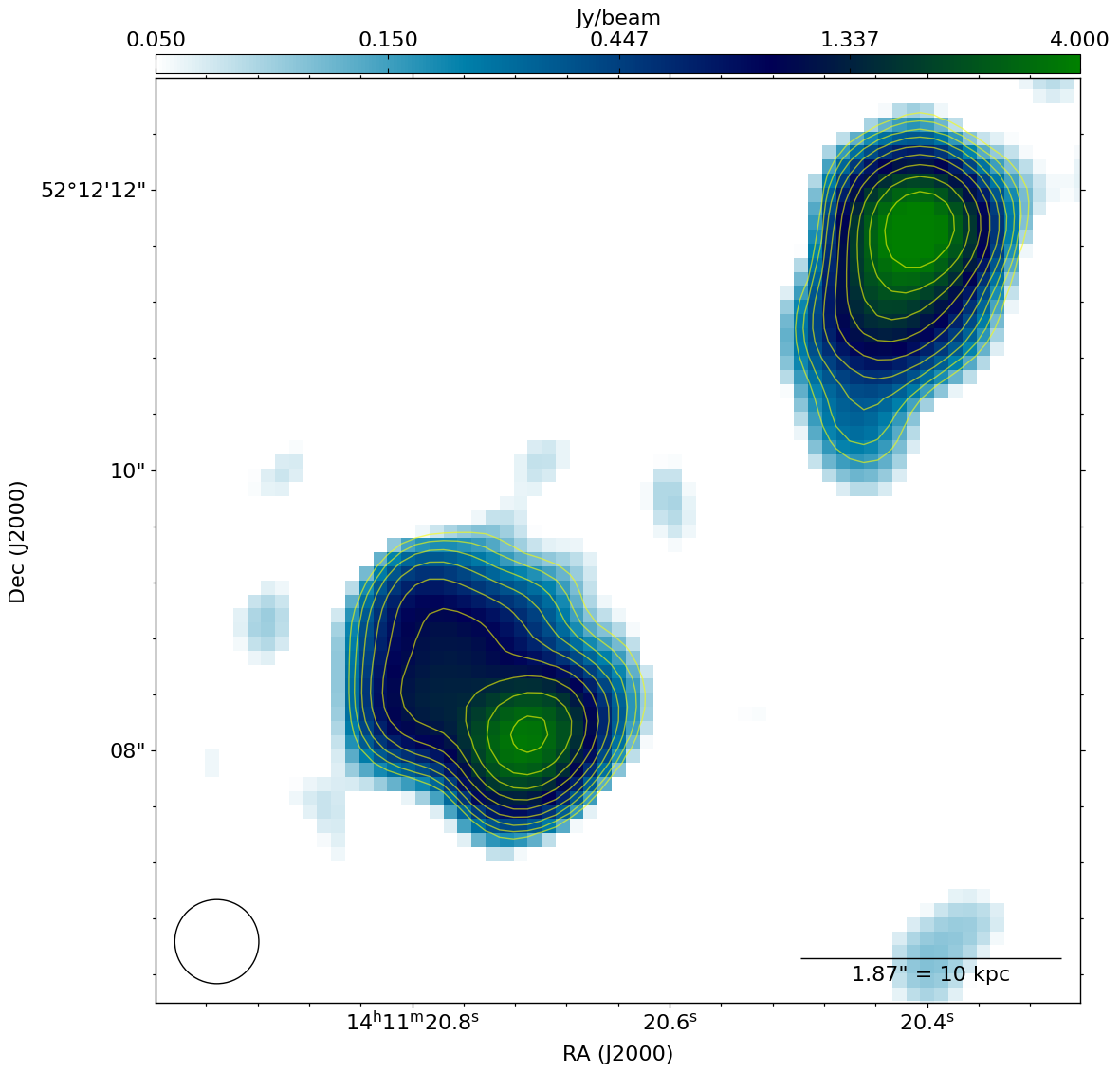}}
		\caption{MERLIN, $1.658$ GHz} \label{fig.imsci.emerlin}
	\end{subfigure}
	\hfill
	\begin{subfigure}{.49\textwidth}
		\resizebox{\hsize}{!}{\includegraphics[width=.32\linewidth]{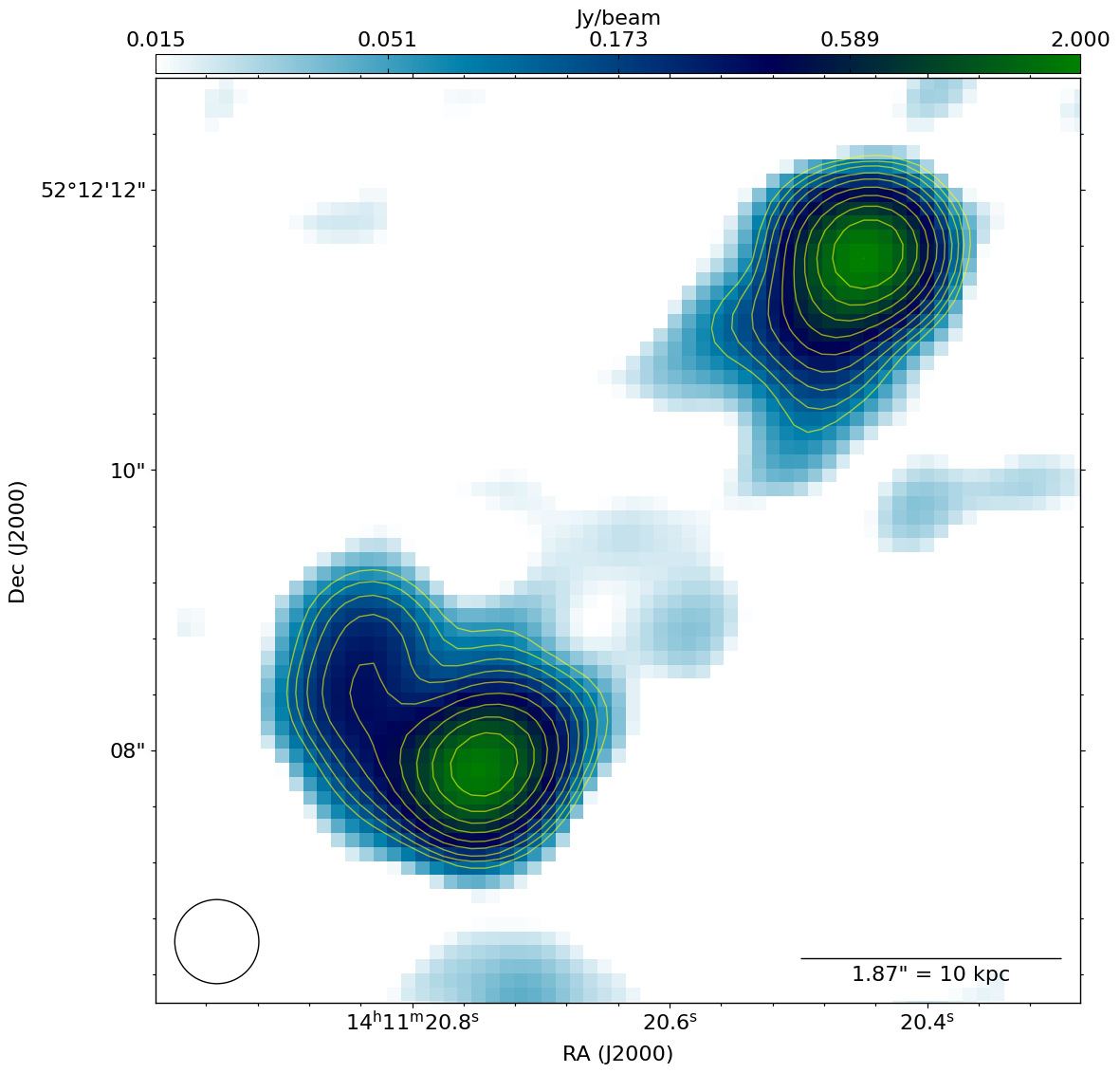}}
		\caption{VLA , $4.760$ GHz} \label{fig.imsci.vla.5ghz}
	\end{subfigure}
	\hfill
	\begin{subfigure}{.49\textwidth}
		\resizebox{\hsize}{!}{\includegraphics[width=.32\linewidth]{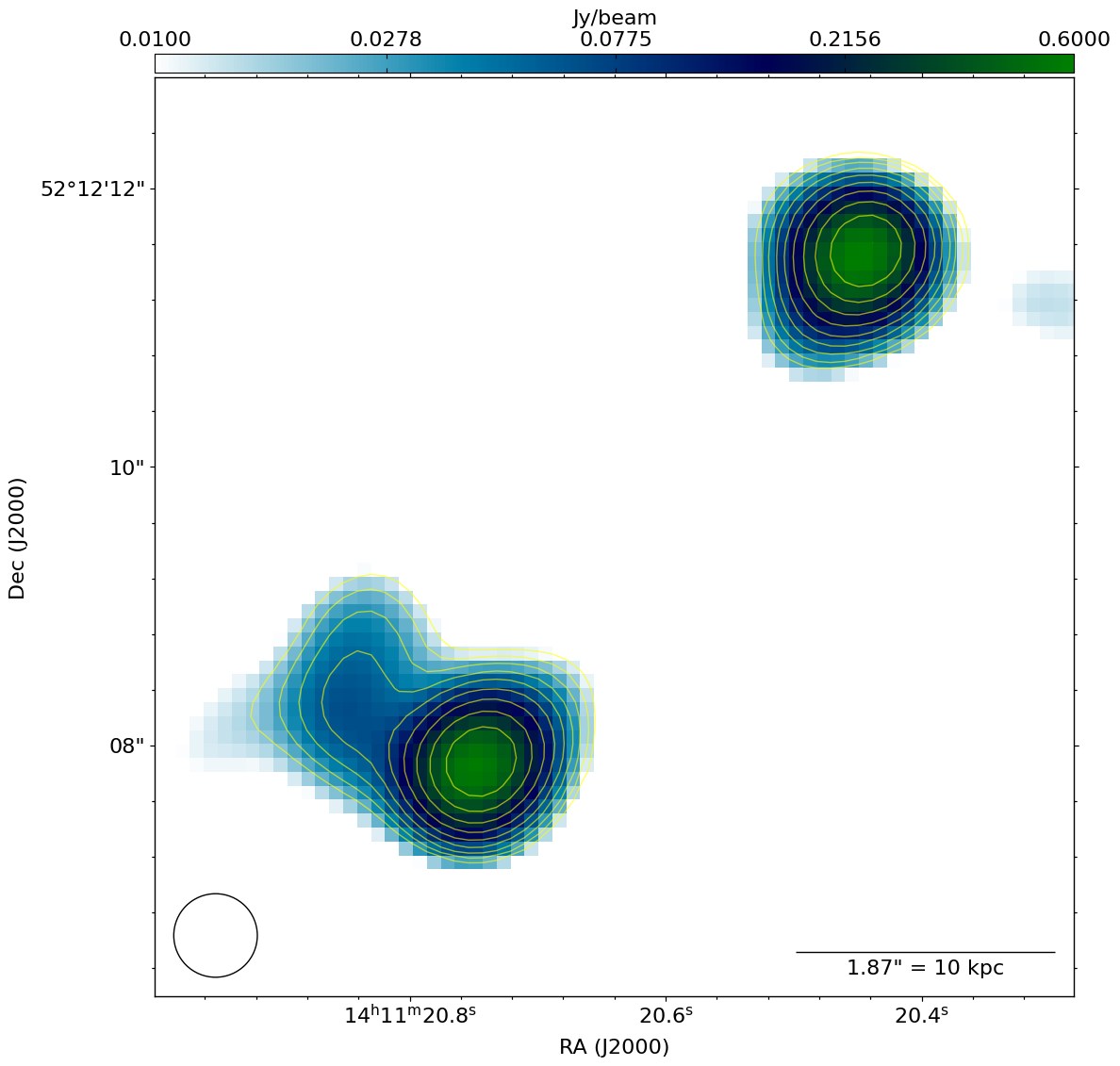}}
		\caption{VLA , $14.975$ GHz} \label{fig.imsci.vla.15ghz}
	\end{subfigure}
	\caption{\label{fig.imsci.fluxes} High resolution images of 3C295 at different frequencies. Flux density units are Jy/bm, with colour scales starting at $5\sigma$ and tuned to show the source morphology. Each image includes an overlay of itself, with 10 levels starting at $3\sigma$ and increasing exponentially until reaching the maximum value in the image. The restoring beam is 0.6$^{\prime\prime}$ in all cases, and the general properties are summarised in Table. \ref{table.ims.recap}.}
\end{figure*}

\pg
In Fig. \ref{fig.imsci.lofar0}, we show our main result: the first image ever produced of 3C295 at 132 MHz with a resolution of $0.4^{\prime\prime}$. Images of the source at higher frequencies, produced using the archival data described previously, are also shown in Fig. \ref{fig.imsci.fluxes}. The properties of these images (name, frequency, noise, dynamic range) are given in Table \ref{table.ims.recap}.
They are all created using similar $uv$-tapers and $uv$-cuts and thus have matching resolutions and $uv$-coverage. They are also spatially aligned on the hotspots. They have all been tied to the flux density scale described in \citet{2017ApJS..230....7P}, since the \citet{2012MNRAS.423L..30S} flux scale used during self-calibration diverges at higher frequencies.The flux scale error is simply taken to be the reference $5\%$ at all frequencies, as 3C295 is one of the calibrator sources in \citet{2017ApJS..230....7P}.

\begin{table}[h!]
	\centering
	\begin{tabular}{cccc}
	\hline
		Image        & Frequency & $\sigma$ & DR      \\
		             & [GHz]     & [mJy/beam] &  \\  \hline \\[-4pt]
		LOFAR        & 0.132     & 0.6 / 30  & 13228\\
		MERLIN       & 1.658     & 26.2 & 189.58   \\
		VLA 5GHz     & 4.760     & 16.4 & 753.42    \\   
		VLA 15GHz    & 14.97     & 2.58 & 216.31     \\ \hline      
	\end{tabular}
	\caption{\label{table.ims.recap} Properties of the images used for our analysis. The images are listed in order of increasing frequency. The dynamic range (DR) is defined as the ratio between the brightest pixel in the image and the thermal noise in the image, far from artefacts. The restoring beam size is $0.6^{\prime\prime}$ in all cases. Where a second $\sigma$ value appears, it is the local noise near the source.}
\end{table}

\pg
Note that, because all these images were made using a Briggs weighting with robust parameter $-2$, which corresponds to uniform weighting, their noise levels are relatively high. The lower colour map thresholds start at the $5\sigma$ level for each image, using the $\sigma$ values listed in Table \ref{table.ims.recap} except for the LOFAR images, which instead use a minimum threshold of 100 mJy/beam. This is due to structure in the residuals near the source. The image is overlaid onto itself as a contour, starting at the $3\sigma$ level.

\subsection{Morphology of 3C295}\label{sec.morphology}

             
\begin{table*}[h]
    \centering
    \begin{tabular}{lccccc}\hline
        Component & $S_{0.132}$ & $S_{1.658}$ & $S_{4.760}$ & $S_{14.975}$  & Offset from host \\
         & [Jy] & [Jy] & [Jy] & [Jy] & [arcsec]\\ \hline \\[-4pt]
        Whole     & 85.63 & 13.89 & 5.25 & 1.48 & 6.1 (LAS) \\
        Hotspot N & 15.53 & 4.82  & 1.97 & 0.56 & 1.92 \\
        Hotspot S & 14.52 & 3.74  & 1.86 & 0.56 & 2.76\\
        Lobe N    & 29.04 & 2.18  & 0.55 & 0.13 & -- \\
        Lobe S    & 26.54 & 3.15  & 0.87 & 0.23 & -- \\ \hline \\
    \end{tabular}
    \caption{
    \label{table.3c295.properties} Table summarising the measured properties of 3C295 and its components, using a 15$\sigma$ flux threshold. For the whole source, we report the Largest Angular Size (LAS) instead of the offset from the host galaxy. Offsets are reported for hotspots only as these were used for aligning the images. }
\end{table*}

\pg
3C295 is a compact source with a largest angular scale of $6.1^{\prime\prime}$, which is fully covered by our shortest baselines at all frequencies. Due to its very high flux density (90.87 Jy at 144 MHz, 19.42 Jy at 1.5 GHz), its features are recovered at high S/N.

\pg
The morphology of 3C295 as observed in our new LOFAR image at 132 MHz agrees well with the morphology observed at higher frequencies: see Fig. \ref{fig.imsci.fluxes} 
\citep[e.g.][]{1991AJ....101.1623P} 
We resolve two separate sources of emission, one in the North-East and one in the South-West, each about 10 kpc wide.

\pg
The Northern component can be divided into sub-components: a dominant compact hot-spot, dominant in flux density, and a fainter lobe surrounding it, which becomes more prominent at lower frequencies. The Northern hotspot is most compact at higher frequencies, and the North lobe emission closest to the host galaxy (situated between the Northern and Southern components) becomes more apparent as frequency decreases. 

\pg
In the Southern component, the hotspot has been reported in the literature as elongated along an East-West axis \citep[e.g.][]{2017CSci..113..707G}. At the resolutions we achieve, we find a high-flux-density, compact source with an extension of more diffuse emission to the North-East. These are referred to, respectively, as the Southern hotspot and Southern lobe. 

\begin{figure}
	\centering
	\includegraphics[width=.99\linewidth]{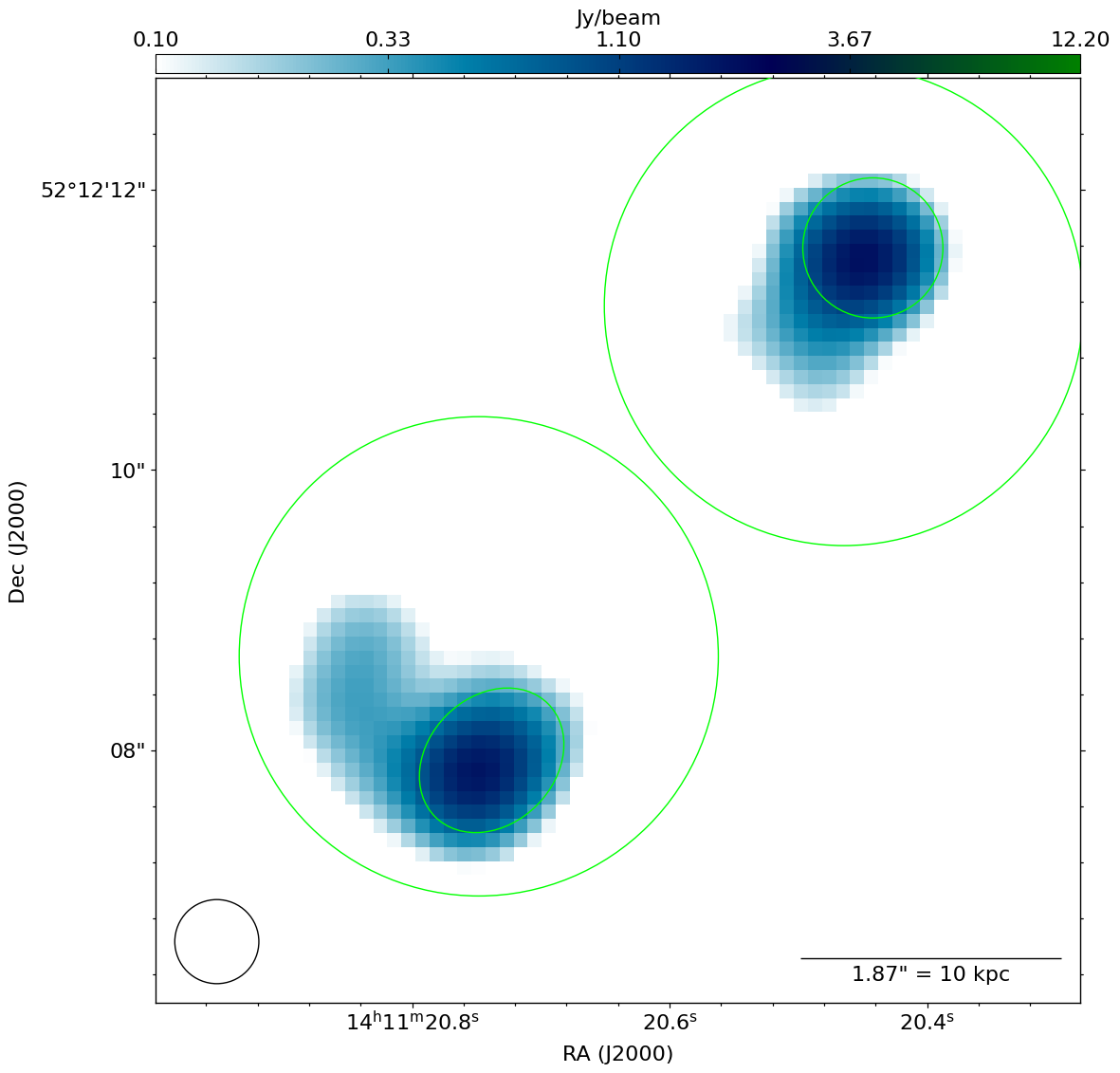}
	\caption{Regions associated with each component: inner ellipses are the "hotspots", outer circles are the "lobes". These are used in combination with a $5\sigma$ flux density threshold, indicated by the overlays in Fig. \ref{fig.imsci.fluxes}} \label{fig.components}
\end{figure}

\pg
These four morphological components are identified with drawn regions in Fig. \ref{fig.components}, and are all larger than the restoring beam ($0.6^{\prime\prime}/3.6$\,kpc).
Although we use only the regions identified in Figure~\ref{fig.components} in this work, the source itself is complex at 132 MHz. 

\section{Resolved Spectral Analysis}\label{sec.spect}

\subsection{Spectral Index Maps \& Radio Colour-Colour Analysis}
    
\begin{figure*}
	\centering
	\begin{subfigure}{.49\textwidth}
		\resizebox{\hsize}{!}{\includegraphics[width=.49\linewidth]{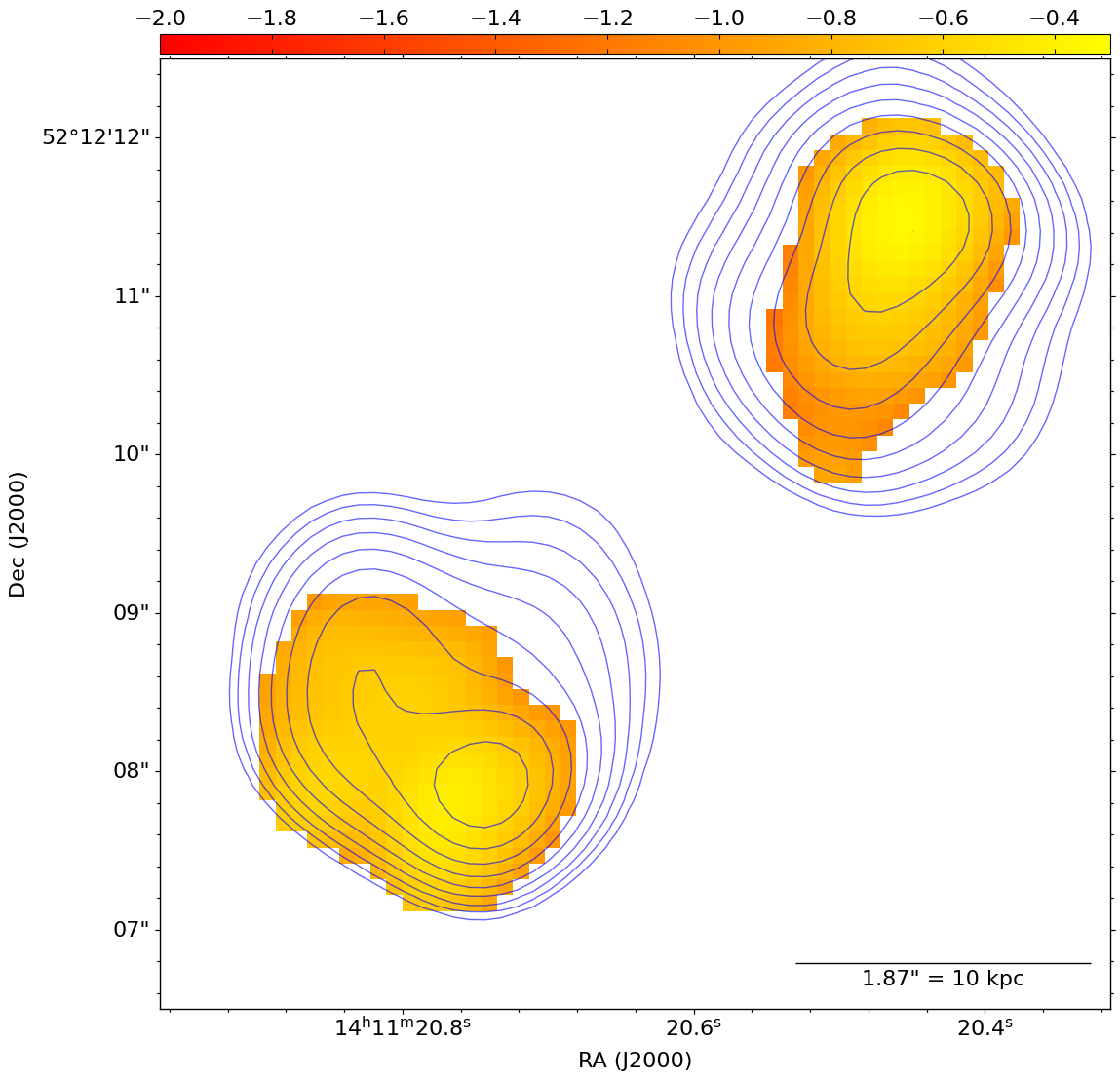}}
		\caption{0.132 - 1.658 GHz spectral indices} \label{fig.spi.lofreqs}
	\end{subfigure}
	\hfill
	\begin{subfigure}{.49\textwidth}
		\resizebox{\hsize}{!}{\includegraphics[width=.49\linewidth]{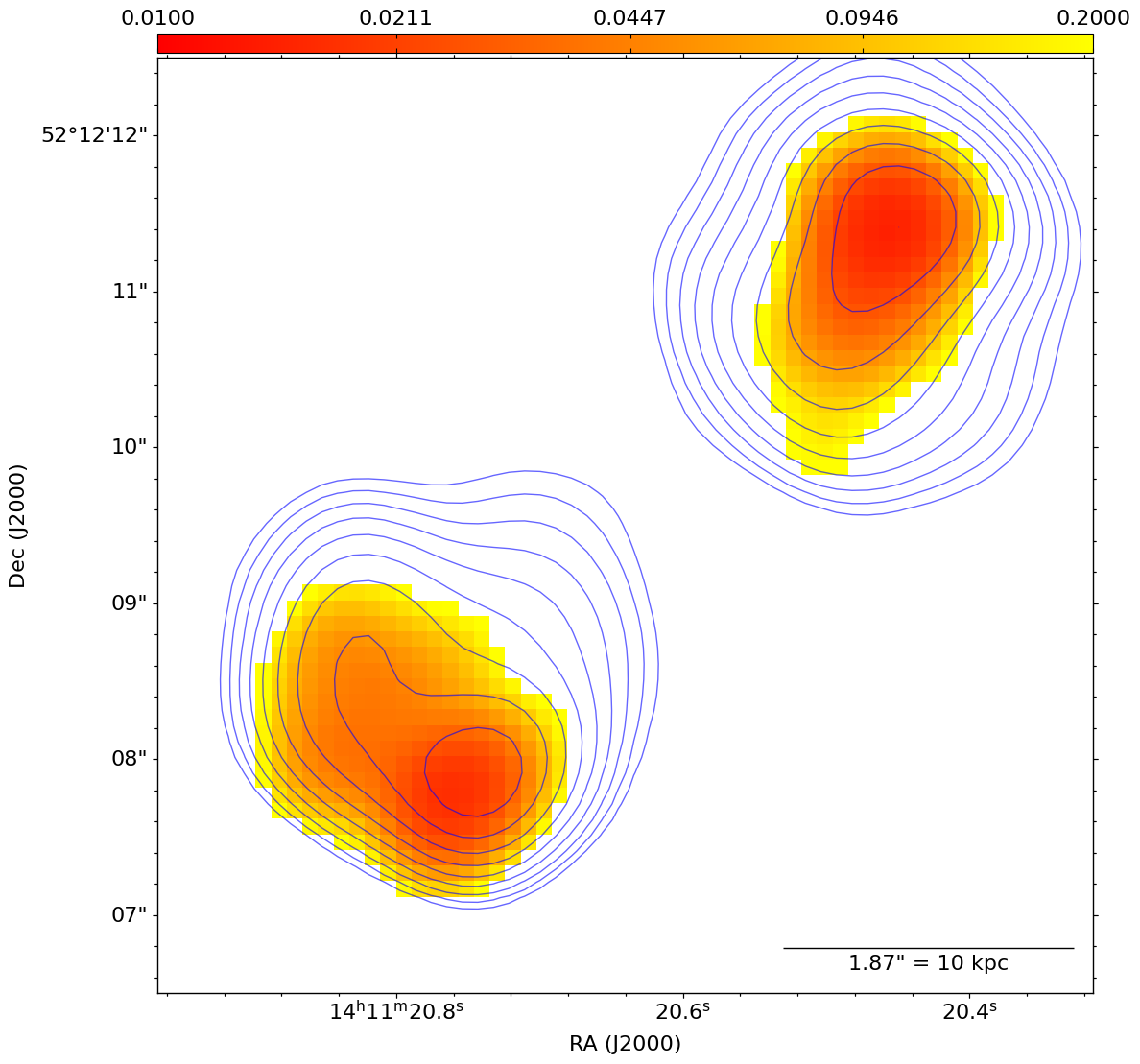}}
		\caption{0.132 - 1.658 GHz spectral index error map} \label{fig.spi.lofreqs.error}
	\end{subfigure}
	\hfill
	\begin{subfigure}{.49\textwidth}
		\resizebox{\hsize}{!}{\includegraphics[width=.49\linewidth]{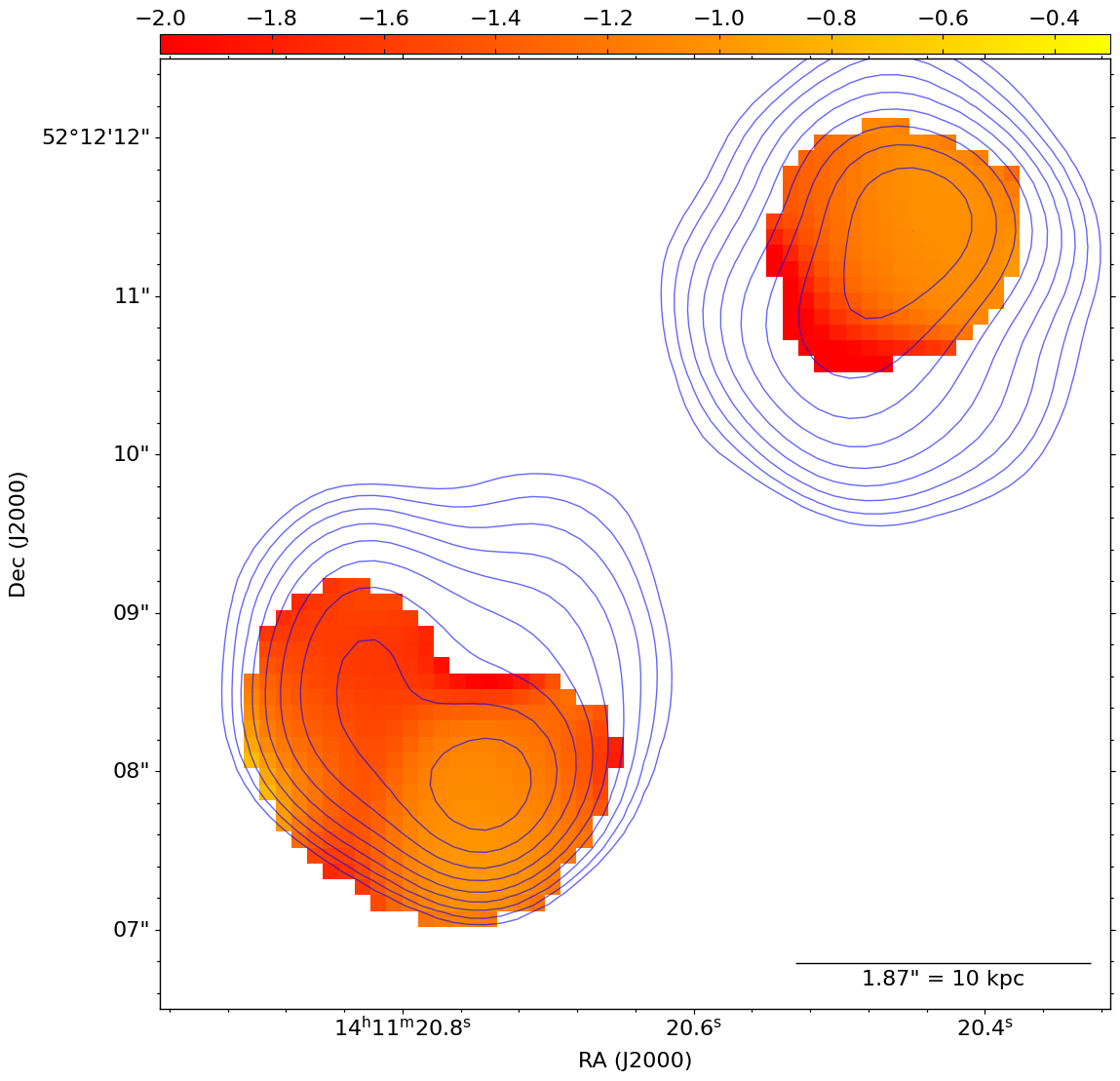}}
		\caption{4.76 - 14.975 GHz spectral indices} \label{fig.spi.vlafreqs}
	\end{subfigure}
	\begin{subfigure}{.49\textwidth}
		\resizebox{\hsize}{!}{\includegraphics[width=.49\linewidth]{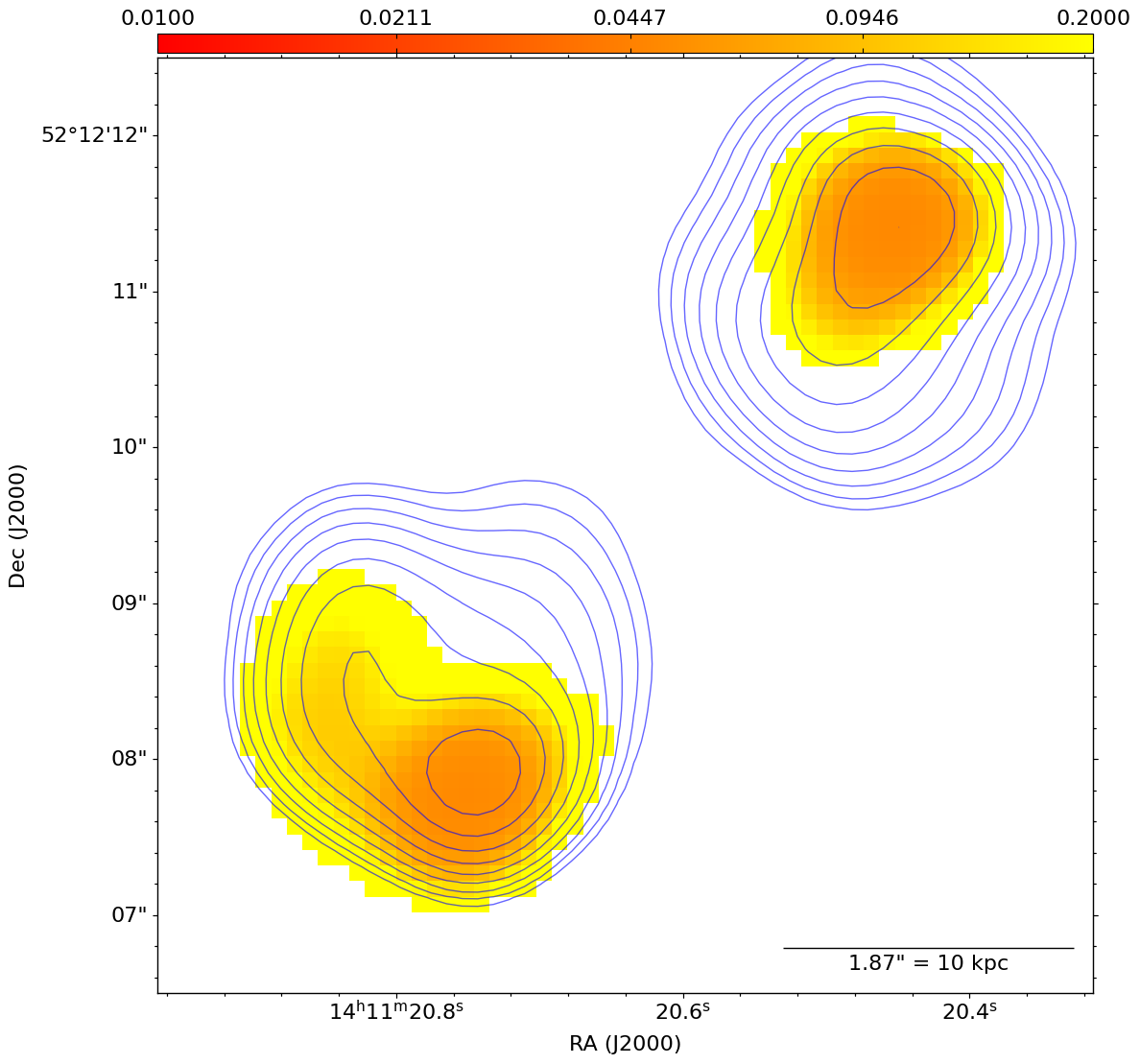}}
		\caption{4.76 - 14.975 GHz spectral index error maps} \label{fig.spi.vlafreqs.error}
	\end{subfigure}
	\hfill
	\caption{ Spectral index maps with associated error maps.}\label{fig.spimaps}
\end{figure*}

\pg
To study the resolved spectral trends within 3C295,
we produce two spectral index maps using the Broadband Radio Astronomy Tools \citep[BRATs][]{2013MNRAS.435.3353H,2015MNRAS.454.3403H}, shown in Fig. \ref{fig.spimaps}: one between 132 MHz and 1.658 GHz, and the other between 4.76 GHz and 14.975 GHz. These are made 
using only pixels with a flux density value above $15\sigma$.

\pg
It is immediately apparent that the overall 0.132 - 1.658 GHz spectral index values are significantly flatter than the 4.76 - 14.975 GHz spectral index values.
The observed spatial distribution trends of the spectral indices are consistent with those typically observed in classical FRII sources \citep{2010A&A...515A..50O,2016MNRAS.463.3143M,2017MNRAS.469..639H}: the flatter indices correspond to the radio hotspots, where particle acceleration takes place, and the overall steepening of the source is in the direction of the host galaxy. In general, we find an overall flattening of the spectral indices towards the lower frequencies, where the spectrum gets closer to the injection value and absorption processes start playing a role.

\pg
We also use these spectral index maps to perform a radio colour-colour analysis \citep{1993ApJ...407..549K}. { These plots show the distribution of spectral indices between one pair of low frequencies ($\alpha_{\rm low}$) versus the spectral indices of another pair of higher frequencies ($\alpha_{\rm high}$). This provides crucial insights into the spectral shape of radio sources. Furthermore, the presence of emission due to mixing of different electron populations and regions with different magnetic field strengths or radiation losses lead to different curves in the colour–colour plot \citep[e.g.][]{2001ASPC..250..372R,Rajpurohit2020a,Rajpurohit2021a}. Radio colour-colour plots are similar to spectral curvature maps - the further a data point lies from the unit line (where $\alpha_{\rm low}=\alpha_{\rm high}$), the more spectral curvature is measured for that point. Similarly, the stronger the curvature, the older the age along a given spectral shape. }

\begin{figure*}
	\centering
    \begin{subfigure}{.49\textwidth}
	\includegraphics[width=.99\linewidth]{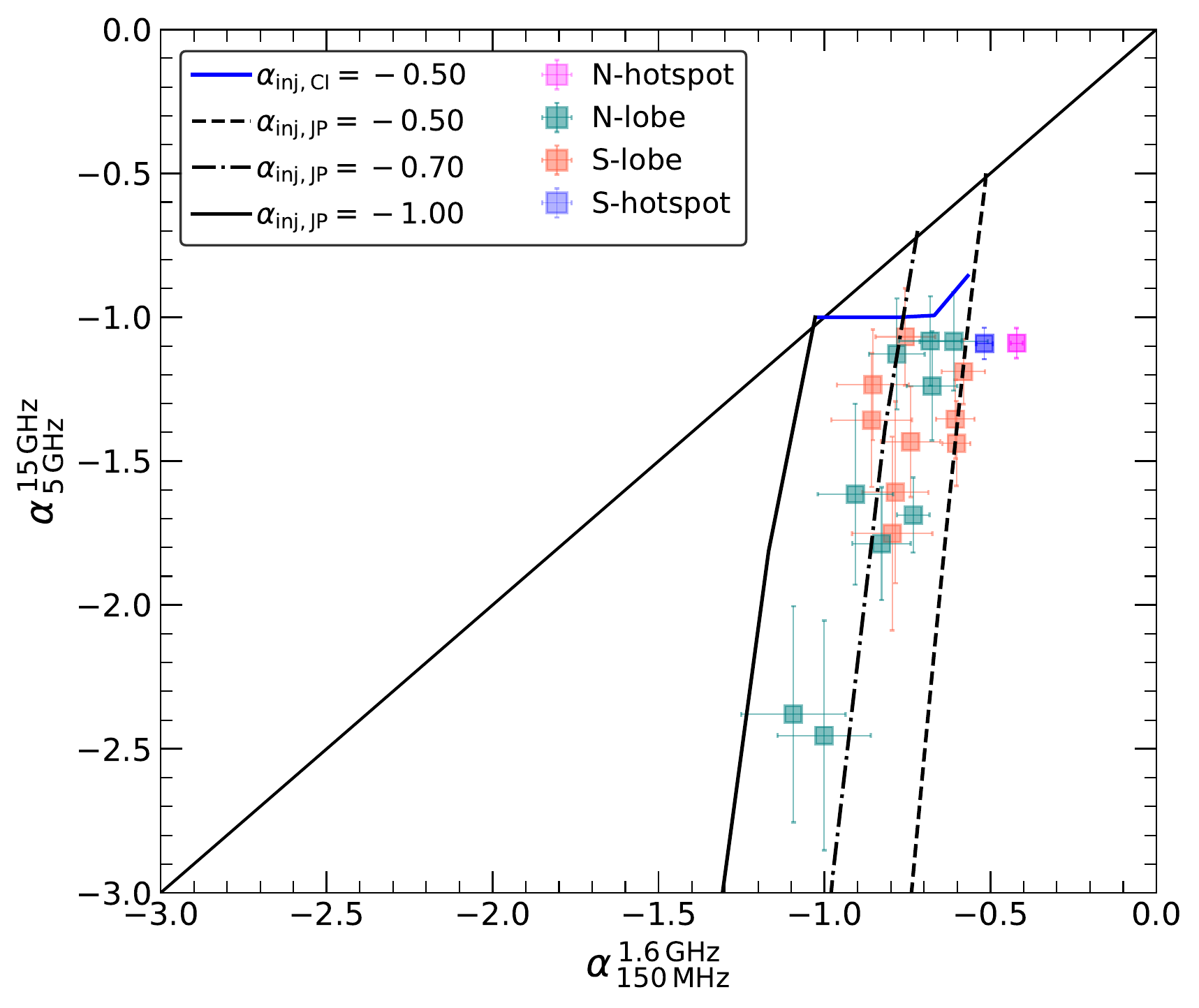}
	\caption{\label{fig.ccplot.diag} Radio colour-colour plot of 3C295 overlaid with the JP and CI spectral aging models with different injection indices. The regions used for extracting the spectral indices are shown in Fig.\ref{fig.ccplot.regs}. The two North lobe outliers correspond to regions near the host galaxy.}
	\end{subfigure}
	\hfill
		\hfill
	\begin{subfigure}{.49\textwidth}
	\includegraphics[width=.99\linewidth]{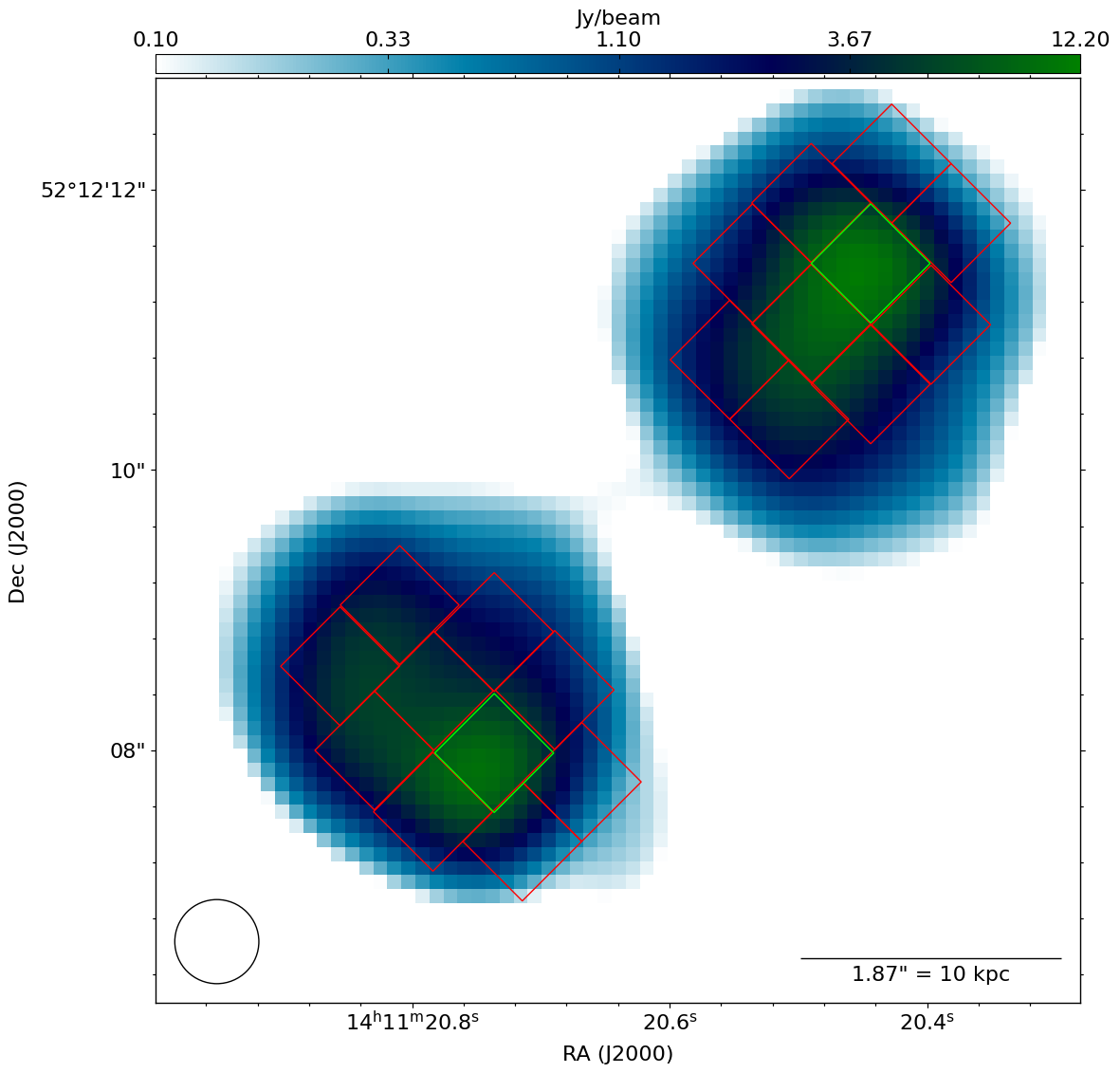}
	\caption{\label{fig.ccplot.regs} Map of the regions used to calculate the colour-colour plot. Red regions are the lobe regions, and green the hotspots.}
	\end{subfigure}
    \begin{subfigure}{.49\textwidth}
	\includegraphics[width=.99\linewidth]{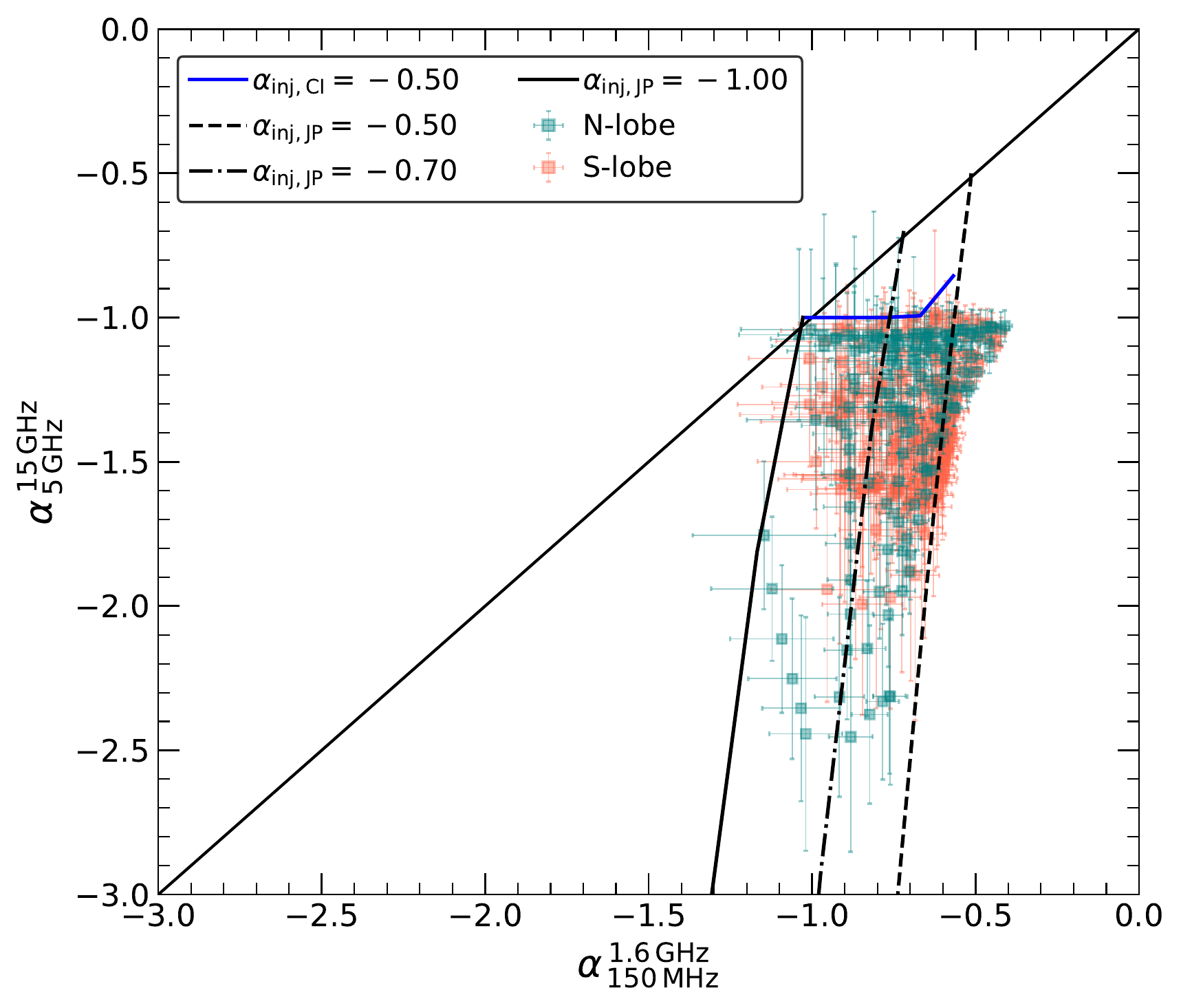}
	\caption{\label{fig.ccplot.pixels.lobes.fit} Per-pixel colour-colour map to the lobes of Fig. \ref{fig.spimaps}. We see that the emission is largely consistent with the JP model. Outliers are pixels which are near the hotspots, and therefore likely to be contaminated by hotspot regions.}
	\end{subfigure}
	\hfill
	\begin{subfigure}{.49\textwidth}
	\includegraphics[width=.99\linewidth]{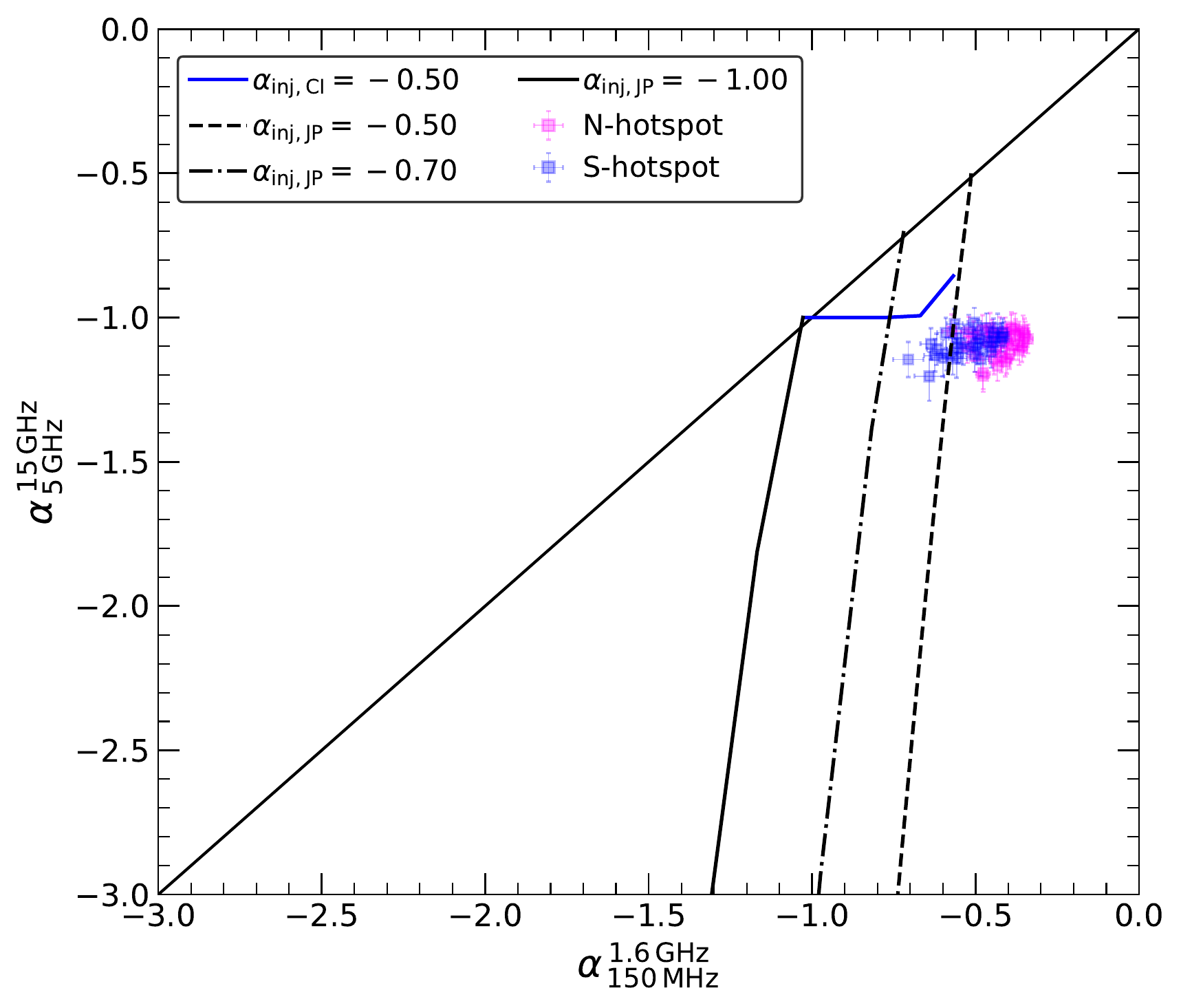}
	\caption{\label{fig.ccplot.pixels.hotspot.fit} Per-pixel colour-colour plot of the hotspot data of Fig. \ref{fig.spimaps}. We see that the emission in hotspots is inconsistent with all three ageing models.}
	\end{subfigure}
	\caption{\label{fig.ccplot} Radio colour-colour plots of different regions of 3C295. The spectral index values of Fig. \ref{fig.spi.lofreqs} are plotted in the x-axis and the values of Fig. \ref{fig.spi.vlafreqs} in the y-axis.}
\label{cc_plots}	
\end{figure*}

{ We first create a colour-colour plot extracting the spectral index from square shaped boxes of width $0.6\arcsec$, which correspond to 3.6\,kpc in physical units. This size was chosen because it is equivalent to the beam size of our observations.} These boxes are shown in Fig \ref{fig.ccplot.regs}. The low frequency spectral index values are extracted between 132\,MHz and 1.658\,GHz and the high frequency one between 4.760\,GHz and 14.975\,GHz. The curvature is negative for a convex spectrum. The resulting plot is shown in Fig. \ref{fig.ccplot.diag}, and the relevant regions are shown in Fig. \ref{fig.ccplot.regs}. All data points lie below the unit line (solid black line, where $\alpha_{0.132\,\rm GHz}^{1.658\,\rm  GHz}=\alpha_{\rm 4.760\,GHz}^{14.975\,\rm GHz}$), which indicates a clear overall negative curvature for 3C295. { All regions seem to follow a single, continuous trend in the colour-colour plot, similar to trends observed for other radio galaxies in the literature \citep[e.g. Figs. 3, 10 and 5 in ][respectively]{1993ApJ...407..549K,2020A&A...638A..29B, 2015A&A...583A..89S}. The two data points furthest from the power-law line - which are therefore associated with the oldest emission in the source - correspond to the South-Westernmost boxes in the North lobe: in other words, with the regions nearest to the host galaxy. } 

\pg
Because we only have access to one box per hotspot, we create a more illustrative pixel-based colour-colour plot, which follows the same trends, albeit more noisily. This allows us to more finely follow the spectral properties of the source.
The resulting plots are shown in Fig. \ref{fig.ccplot.pixels.lobes.fit} and \ref{fig.ccplot.pixels.hotspot.fit}. It is evident that the spectral shape of the Northern and Southern hotspots are different than the Northern and Southern lobes. 

\pg
To better understand the curvature distribution within the individual components of 3C295, we superimposed the observed data with standard spectral ageing models.  { Indeed, radio colour-colour plots have the advantage that the standard spectral models depend only on the injection index, and are independent of magnetic field strength, adiabatic expansion/compression or radiative losses  \citep{1993ApJ...407..549K,vanWeeren2012a}. They represent an easy way to visualise the ageing models and to trace back the data to injection properties. The extrapolation of the trajectory in the colour-colour plot to the power-law line allows us to estimate the injection index of radio source.}

\pg
We consider two standard spectral models. In the Jaffe-Perola (JP) spectral model \citep{1973A&A....26..423J}, the electron population of a given region is accelerated once, resulting in a power-law energy distribution and therefore a power-law flux density profile. This model also assumes that electrons undergo a series of scattering events which randomise their pitch angle relative to the magnetic field lines. The continuous injection (CI) model \citep{1970ranp.book.....P} assumes instead, as the name implies, ongoing injection of fresh particles into a given region. 

\pg
{As a reference, we overlay four curves to the data, which represent JP models with injection indices equal to $-0.50$, $-0.70$, and $-1.0$, and a CI model with injection index equal to $-0.50$.}
{ The $-0.50$ is the minimum injection index commonly used as a lower bound and the other two values demonstrate how the curve would shift in the plot when increasing the injection index. We note that most of the points related to the source's lobe  in Figs.\,\ref{fig.ccplot.diag} and \ref{fig.ccplot.pixels.lobes.fit} can be described with a JP model with injection index of $\sim-0.70$. 

Interestingly, the data points associated with 3C295's hotspots have a significantly different distribution from the lobes (Fig. \ref{fig.ccplot.pixels.hotspot.fit}) with values varying mainly in the range of $-0.3<\alpha<-0.5$. This distribution is inconsistent with the JP and CI models. Such flat spectral indices could be indicative of synchrotron self-absorption, thermal absorption, or a low-energy cut-off \citep[e.g.][]{1997ApJ...479..258K}; in this latter case, given the magnetic field of the hotspots constrained by equipartition \& SSC models \citep{2000ApJ...530L..81H}, a low-energy cutoff or flattening in the hotspot electron spectrum would be present at $\gamma \sim 300$. The hotspot data distribution in the colour-colour plot can thus best be understood as evidence of one or more of these low-frequency absorption processes taking place within the hotspots.}

\pg
{It is worth mentioning that some points related to the lobes shown in \ref{fig.ccplot.pixels.lobes.fit} appear to be located in the same area of the plot where the hotspot points lie (on the left side of the JP model with injection index -0.5). These points are likely coming from regions where there is an overlap between the lobe and hotspot flux density contribution.}

\pg

{In conclusion, the spectral trend in the lobes is consistent with a JP model, indicative of particle radiative cooling. To the contrary, hotspot data provides evidence for the presence of absorption processes, as we will further investigate in the next section.}

\subsection{Component Spectral Analysis}

{
\pg
To further investigate the signs of absorption in the low-frequency spectra of the hotspots identified in the previous section, we perform a spectral fit to the radio spectral energy distributions of the hotspots. Spectral modelling is often done using BRATs, but this software package has not yet implemented low-frequency absorption models. We thus fit our hotspot radio spectral energy distributions using two classical absorption models: synchrotron self-absorption and free-free absorption.
}

\pg
Using the morphological components defined in Fig. \ref{fig.components}, we integrate the flux density in our four morphological regions, subtracting the hotspot region flux density from the total lobe region flux density for both North and South. We assume that the underlying spectrum before absorption does not change over time or within a given region, which is necessary to reduce the number of degrees of freedom to fewer than the number of data points for all our absorption models - synchrotron self-absorption (SSA), free-free absorption (FFA), and power law (PL) spectra. 
The latter is only used as a point of comparison.

\pg
Here we briefly outline the spectral models used, given in Eqns. \ref{eq.PL}-\ref{eq.SSA} below. Further details on these models can be found in e.g. \citet{2015ApJ...809..168C}.

For the power-law spectral model, given by

\begin{equation}
S_\text{PL} (S_0,\alpha_\text{PL}) = S_0 \left(\frac{\nu}{\nu_0}\right)^{\alpha_\text{PL}},   \label{eq.PL}
\end{equation}

there are two free parameters: a flux density normalisation $S_0$ (corresponding to the flux density of the source at 150 \,MHz in our case) and the spectral index $(\alpha_\text{PL})$. It is therefore well-constrained by our data. From Fig. \ref{fig.ccplot}, we expect this model to be a poor fit for all components - no part of our source lies on the unit line of our colour-colour plot. It is therefore used as a point of comparison and contrast with the other models.

The free-free absorption model, given by:

\begin{equation}
S_\text{FFA} (S_0,\nu_c, \alpha_\text{FFA})  = S_0 \left(\frac{\nu}{\nu_0}\right)^{\alpha_\text{FFA}} \exp \left[{- \left(\frac{\nu}{\nu_c}\right)^{-2.1}}\right],   \label{eq.FFA}  \\
\end{equation}
has 3 free parameters: flux density normalisation $S_0$, spectral index $\alpha_\text{FFA}$, and the absorption frequency $\nu_c$, which is the frequency at which the optical depth is unity. This model is also well-constrained by our available data.

Finally, for the synchrotron self-absorption model we have
\begin{align}
S_\text{SSA} (S_0,\nu_c, \beta)  &= S_0 \left( \frac{\nu}{\nu_c}\right)^\frac{\beta-1}{2} \left( \frac{1-e^{\tau}}{\tau} \right) \label{eq.SSA}  \\
\text{where}\quad\tau&= \left( \frac{\nu}{\nu_c}\right)^\frac{-\left( \beta + 4\right)}{2}. \notag
\end{align}
\pg
Eq. \ref{eq.SSA} describes synchrotron self-absorption with 3 free parameters:  $S_0$, the flux density at $150$\,MHz, $\nu_c$, the absorption frequency, which is the frequency at which the optical depth is unity, and $\beta$, the power law index of the energy distribution of the emitting electron population. $\beta$ is related to the synchrotron spectral index as $\alpha_\text{SSA} = \left(\beta-1\right)/2$.

\begin{figure*}
	\centering
	\includegraphics[height=.8\linewidth]{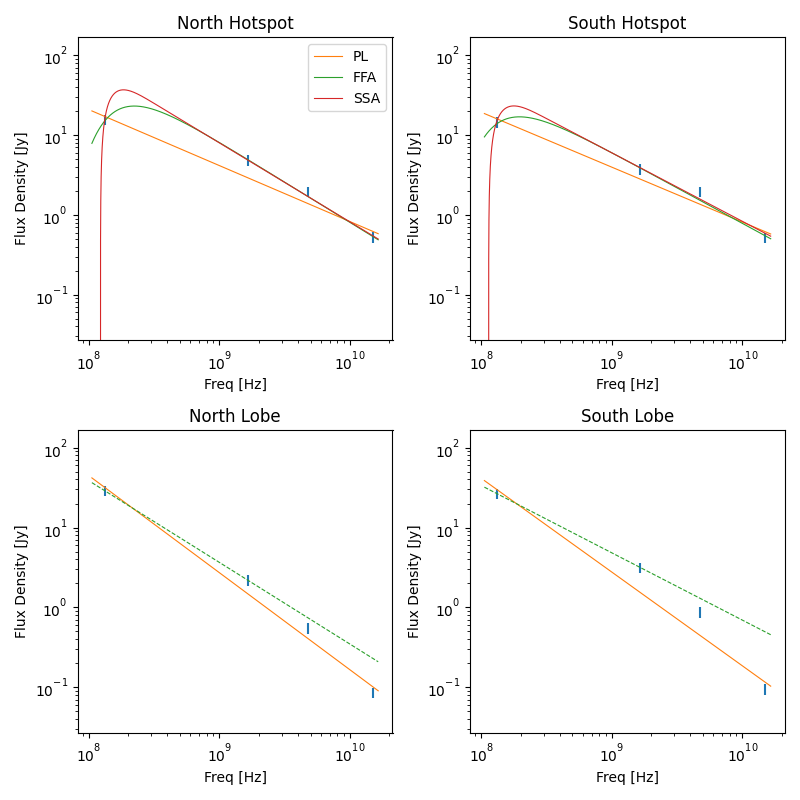}
	\caption{\label{fig.spifit} Per-component spectral fitting, using synchrotron self-absorption, free-free absorption and simple power law models. The model colours are the same in all plots. The dashed line in the lobe plots represents a power-law fit done using only the two lowest-frequency data points.}
\end{figure*}

\pg
We perform fits using scipy's \texttt{curve\_fit} function, which performs a Levenberg-Marquardt fit between our function and our data points. These fits are then used as the initial parameters of our Markov-chain Monte-Carlo (MCMC) fitting method. We employed an `affine invariant' MCMC ensemble sampler \citep{Goodman2010} as implemented by the \textsc{emcee} package \citep{emcee}, creating 500 `walkers' to fit the data, with the $1\sigma$ confidence interval for each of our fits used as a final error estimate. 
The points and the best-fit curves are all shown in Fig. \ref{fig.spifit}. 
The best-fit parameter values and their errors are summarised in Table \ref{tab.fitvals}.

\begin{table*}
\centering
\begin{tabular}{c|c|c|c|c}
                         & N Hotspot      & N Lobe         & S Hotspot      & S Lobe \\\hline
 PL                      &                &                &                &                 \\
     $S_0$ [Jy]          & $4.1\pm0.4$    & $2.7\pm0.2$    & $4.0\pm0.3$    & $2.77\pm0.2$     \\
     $\alpha_\text{PL}$  & $-0.70\pm0.04$ & $-1.22\pm0.04$ & $-0.69\pm0.03$ & $-1.17\pm0.03$  \\
     $\chi^2_r$          & 6.38           &  4.8           & 4.44           & 12.6            \\\hline
FFA                      &                &                &                &                 \\
     $S_0$ [Jy]          & $8.2\pm1.5$    &                & $6.1\pm1.2$    &                 \\
     $\nu_c$ [MHz]       & $158\pm19$     &                & $130\pm23$     &                 \\
     $\alpha_\text{FFA}$ & $-1.01\pm0.1$  &                & $-0.89\pm0.08$ &                 \\
     $\chi^2_r$          &  0.31          &                & 0.83           &                 \\\hline
SSA                      &                &                &                &                 \\
     $S_0$ [Jy]          & $50.8\pm18.0$  &                & $28.9\pm10.3$  &                 \\
     $\nu_c$ [MHz]       & $154\pm3$      &                & $143\pm 10$    &                 \\
     $\alpha_\text{SSA}$ & $-0.99\pm0.41$ &                & $-0.83\pm0.39$ &                 \\
     $\chi^2_r$          & 0.3            &                & 0.81           &                 
\end{tabular}
\caption{\label{tab.fitvals} Best-fit parameters \& associated errors for all three spectral models fitted. $\chi^2_r$ are the reduced $\chi^2$ values for each fit.}
\end{table*}


\pg
Fig. \ref{fig.spifit} shows the spectral behaviour of our 4 components along with the spectral model best-fit curves. We see that the SSA and FFA models both seem give good fits to the data. However, only the LOFAR data point lies below the turnover frequency, which means that we lack the low-frequency constraints to discriminate between FFA and SSA.

\pg
We also see that the power law fit fails to describe the low-frequency properties of the hotspots, or the high-frequency properties of the lobes. This is also apparent in the power-law fit $\chi^2_r$ values, which are systematically higher than 1 - the model is not complex enough to capture the spectral behaviour of the data.

\pg
For all components, the FFA and SSA spectra are in very close agreement at high frequencies. 
The flux normalisation at 150 \,MHz varies between the fits for all components, but this seems to be strongly tied to the un-physically sharp decrease in the best-fit SSA spectra at lower frequencies (see Fig. \ref{fig.spifit}). An additional ILT measurement at 55 MHz
will allow further resolved spectral analysis of 3C295. This will be more challenging due to the lower resolution that comes with lower frequencies, but it will help that the integrated spectrum of 3C295 peaks at about 50 MHz\citep{2012MNRAS.423L..30S}.

%% file: source/conclusion.tex
\section{Conclusion \& Future Work}\label{sec.conclusion}

\pg
In this paper, we have used the sub-arcsecond capabilities of the ILT to create a spatially resolved map of 3C295 at 132 MHz. Combined with archival data from the VLA and MERLIN at GHz frequencies, we analyse four components: the Northern and Southern hotspots and two lobes. While a low-frequency turnover had previously been identified in the integrated spectra for 3C295, here we clearly show that this is due to low-frequency absorption processes in the hotspots. The lobes remain consistent with un-absorbed, optically-thin synchrotron emission, which { can be described by a JP particle ageing model.}

\pg
While the hotspot spectral measurements are consistent with the presence of the low-frequency absorption, we are unable to distinguish between synchrotron self-absorption and free-free absorption. Doing so will require independent observations at lower frequencies. ILT Low Band Antenna (LBA) observations at 55 MHz would provide a strong constraint on the low frequency spectral behaviour, but currently the degradation in resolution by a factor of 2 when compared to the HBA means we cannot differentiate between the lobe and hotspot emission. The ILT is expected to expand in the coming years, notably with the recent addition of a Latvian station to the array, and the immenent addition of an Italian station. These will further improve $uv$-coverage and resolution of the ILT.

\pg
Finally, in order to maximise our scientific returns, the following advances must in our estimation be made, listed in order of expected difficulty and significance:
\begin{itemize}
    \item Formalise the limitations imposed by the pixel statistics in interferometric images for spectral analysis,
    \item Formalise the limitations imposed by joint deconvolution processes in very-large-bandwidth analysis,
    \item Formalise the limitations imposed by the conformity of $uv$-coverage across multi-frequency observations.
\end{itemize}


\pg
Sub-arcsecond imaging with the ILT is paving the way for spatially resolved spectral modelling of compact radio-loud AGN like 3C295. Future implementation of low-frequency absorption models to BRATs and elsewhere     will allow physically motivated modelling that already takes into account the above list of limitations. The continuation of post-processing LoTSS at sub-arcsecond resolution (Morabito et al., 2021), will allow us to perform this kind of spectral modelling for large numbers of sources in the near future.

\pg

%% file: paper.bbl
\begin{thebibliography}{45}
\expandafter\ifx\csname natexlab\endcsname\relax\def\natexlab#1{#1}\fi

\bibitem[{{Ahn} \& {et al.}(2013)}]{2013yCat.5139....0A}
{Ahn}, C.~P. \& {et al.} 2013, VizieR Online Data Catalog, V/139

\bibitem[{{Akujor} {et~al.}(1990){Akujor}, {Spencer}, \&
  {Wilkinson}}]{1990MNRAS.244..362A}
{Akujor}, C.~E., {Spencer}, R.~E., \& {Wilkinson}, P.~N. 1990, \mnras, 244, 362

\bibitem[{{Bonnassieux} {et~al.}(2018){Bonnassieux}, {Tasse}, {Smirnov}, \&
  {Zarka}}]{2018A&A...615A..66B}
{Bonnassieux}, E., {Tasse}, C., {Smirnov}, O., \& {Zarka}, P. 2018, \aap, 615,
  A66

\bibitem[{{Brienza} {et~al.}(2020){Brienza}, {Morganti}, {Harwood}, {Duchet},
  {Rajpurohit}, {Shulevski}, {Hardcastle}, {Mahatma}, {Godfrey}, {Prandoni},
  {Shimwell}, \& {Intema}}]{2020A&A...638A..29B}
{Brienza}, M., {Morganti}, R., {Harwood}, J., {et~al.} 2020, \aap, 638, A29

\bibitem[{{Briggs}(1995)}]{1995AAS...18711202B}
{Briggs}, D.~S. 1995, in Bulletin of the American Astronomical Society,
  Vol.~27, American Astronomical Society Meeting Abstracts, 1444

\bibitem[{{Brunetti} {et~al.}(2001){Brunetti}, {Cappi}, {Setti}, {Feretti}, \&
  {Harris}}]{2001A&A...372..755B}
{Brunetti}, G., {Cappi}, M., {Setti}, G., {Feretti}, L., \& {Harris}, D.~E.
  2001, \aap, 372, 755

\bibitem[{{Callingham} {et~al.}(2015){Callingham}, {Gaensler}, {Ekers},
  {Tingay}, {Wayth}, {Morgan}, {Bernardi}, {Bell}, {Bhat}, {Bowman}, {Briggs},
  {Cappallo}, {Deshpande}, {Ewall-Wice}, {Feng}, {Greenhill}, {Hazelton},
  {Hindson}, {Hurley-Walker}, {Jacobs}, {Johnston-Hollitt}, {Kaplan},
  {Kudrayvtseva}, {Lenc}, {Lonsdale}, {McKinley}, {McWhirter}, {Mitchell},
  {Morales}, {Morgan}, {Oberoi}, {Offringa}, {Ord}, {Pindor}, {Prabu},
  {Procopio}, {Riding}, {Srivani}, {Subrahmanyan}, {Udaya Shankar}, {Webster},
  {Williams}, \& {Williams}}]{2015ApJ...809..168C}
{Callingham}, J.~R., {Gaensler}, B.~M., {Ekers}, R.~D., {et~al.} 2015, \apj,
  809, 168

\bibitem[{{de Gasperin} {et~al.}(2019){de Gasperin}, {Dijkema}, {Drabent},
  {Mevius}, {Rafferty}, {van Weeren}, {Br{\"u}ggen}, {Callingham}, {Emig},
  {Heald}, {Intema}, {Morabito}, {Offringa}, {Oonk}, {Orr{\`u}},
  {R{\"o}ttgering}, {Sabater}, {Shimwell}, {Shulevski}, \&
  {Williams}}]{2019A&A...622A...5D}
{de Gasperin}, F., {Dijkema}, T.~J., {Drabent}, A., {et~al.} 2019, \aap, 622,
  A5

\bibitem[{{Edge} {et~al.}(1959){Edge}, {Shakeshaft}, {McAdam}, {Baldwin}, \&
  {Archer}}]{1959MmRAS..68...37E}
{Edge}, D.~O., {Shakeshaft}, J.~R., {McAdam}, W.~B., {Baldwin}, J.~E., \&
  {Archer}, S. 1959, \memras, 68, 37

\bibitem[{{Foreman-Mackey} {et~al.}(2013){Foreman-Mackey}, {Hogg}, {Lang}, \&
  {Goodman}}]{emcee}
{Foreman-Mackey}, D., {Hogg}, D.~W., {Lang}, D., \& {Goodman}, J. 2013,
  Publications of the Astronomical Society of the Pacific, 125, 306

\bibitem[{{Gilbert} {et~al.}(2004){Gilbert}, {Riley}, {Hardcastle}, {Croston},
  {Pooley}, \& {Alexander}}]{2004MNRAS.351..845G}
{Gilbert}, G.~M., {Riley}, J.~M., {Hardcastle}, M.~J., {et~al.} 2004, \mnras,
  351, 845

\bibitem[{{Goodman} \& {Weare}(2010)}]{Goodman2010}
{Goodman}, J. \& {Weare}, J. 2010, Communications in Applied Mathematics and
  Computational Science, Vol. 5, No. 1, p. 65-80, 2010, 5, 65

\bibitem[{{Gupta} {et~al.}(2017){Gupta}, {Ajithkumar}, {Kale}, {Nayak},
  {Sabhapathy}, {Sureshkumar}, {Swami}, {Chengalur}, {Ghosh},
  {Ishwara-Chandra}, {Joshi}, {Kanekar}, {Lal}, \& {Roy}}]{2017CSci..113..707G}
{Gupta}, Y., {Ajithkumar}, B., {Kale}, H.~S., {et~al.} 2017, Current Science,
  113, 707

\bibitem[{{Harris} {et~al.}(2000){Harris}, {Nulsen}, {Ponman}, {Bautz},
  {Cameron}, {David}, {Donnelly}, {Forman}, {Grego}, {Hardcastle}, {Henry},
  {Jones}, {Leahy}, {Markevitch}, {Martel}, {McNamara}, {Mazzotta}, {Tucker},
  {Virani}, \& {Vrtilek}}]{2000ApJ...530L..81H}
{Harris}, D.~E., {Nulsen}, P.~E.~J., {Ponman}, T.~J., {et~al.} 2000, \apjl,
  530, L81

\bibitem[{{Harwood} {et~al.}(2015){Harwood}, {Hardcastle}, \&
  {Croston}}]{2015MNRAS.454.3403H}
{Harwood}, J.~J., {Hardcastle}, M.~J., \& {Croston}, J.~H. 2015, \mnras, 454,
  3403

\bibitem[{{Harwood} {et~al.}(2013){Harwood}, {Hardcastle}, {Croston}, \&
  {Goodger}}]{2013MNRAS.435.3353H}
{Harwood}, J.~J., {Hardcastle}, M.~J., {Croston}, J.~H., \& {Goodger}, J.~L.
  2013, \mnras, 435, 3353

\bibitem[{{Harwood} {et~al.}(2017){Harwood}, {Hardcastle}, {Morganti},
  {Croston}, {Br{\"u}ggen}, {Brunetti}, {R{\"o}ttgering}, {Shulevski}, \&
  {White}}]{2017MNRAS.469..639H}
{Harwood}, J.~J., {Hardcastle}, M.~J., {Morganti}, R., {et~al.} 2017, \mnras,
  469, 639

\bibitem[{{Jaffe} \& {Perola}(1973)}]{1973A&A....26..423J}
{Jaffe}, W.~J. \& {Perola}, G.~C. 1973, \aap, 26, 423

\bibitem[{{Katz-Stone} \& {Rudnick}(1997)}]{1997ApJ...479..258K}
{Katz-Stone}, D.~M. \& {Rudnick}, L. 1997, \apj, 479, 258

\bibitem[{{Katz-Stone} {et~al.}(1993){Katz-Stone}, {Rudnick}, \&
  {Anderson}}]{1993ApJ...407..549K}
{Katz-Stone}, D.~M., {Rudnick}, L., \& {Anderson}, M.~C. 1993, \apj, 407, 549

\bibitem[{{Mathieu} \& {Spinrad}(1981)}]{1981ApJ...251..485M}
{Mathieu}, R.~D. \& {Spinrad}, H. 1981, \apj, 251, 485

\bibitem[{{McKean} {et~al.}(2016){McKean}, {Godfrey}, {Vegetti}, {Wise},
  {Morganti}, {Hardcastle}, {Rafferty}, {Anderson}, {Avruch}, {Beck}, {Bell},
  {van Bemmel}, {Bentum}, {Bernardi}, {Best}, {Blaauw}, {Bonafede},
  {Breitling}, {Broderick}, {Br{\"u}ggen}, {Cerrigone}, {Ciardi}, {de
  Gasperin}, {Deller}, {Duscha}, {Engels}, {Falcke}, {Fallows}, {Frieswijk},
  {Garrett}, {Grie{\ss}meier}, {van Haarlem}, {Heald}, {Hoeft}, {Horst},
  {Iacobelli}, {Intema}, {Juette}, {Karastergiou}, {Kondratiev}, {Koopmans},
  {Kuniyoshi}, {Kuper}, {van Leeuwen}, {Maat}, {Mann}, {Markoff}, {McFadden},
  {McKay-Bukowski}, {Mulcahy}, {Munk}, {Nelles}, {Orru}, {Paas},
  {Pandey-Pommier}, {Pietka}, {Pizzo}, {Polatidis}, {Reich}, {R{\"o}ttgering},
  {Rowlinson}, {Scaife}, {Serylak}, {Shulevski}, {Sluman}, {Smirnov},
  {Steinmetz}, {Stewart}, {Swinbank}, {Tagger}, {Thoudam}, {Toribio},
  {Vermeulen}, {Vocks}, {van Weeren}, {Wucknitz}, {Yatawatta}, \&
  {Zarka}}]{2016MNRAS.463.3143M}
{McKean}, J.~P., {Godfrey}, L.~E.~H., {Vegetti}, S., {et~al.} 2016, \mnras,
  463, 3143

\bibitem[{{Mohan} \& {Rafferty}(2015)}]{2015ascl.soft02007M}
{Mohan}, N. \& {Rafferty}, D. 2015, {PyBDSF: Python Blob Detection and Source
  Finder}

\bibitem[{{Napier} {et~al.}(1983){Napier}, {Thompson}, \&
  {Ekers}}]{1983IEEEP..71.1295N}
{Napier}, P.~J., {Thompson}, A.~R., \& {Ekers}, R.~D. 1983, IEEE Proceedings,
  71, 1295

\bibitem[{{Offringa}(2010)}]{2010ascl.soft10017O}
{Offringa}, A.~R. 2010, {AOFlagger: RFI Software}

\bibitem[{{Offringa} {et~al.}(2014){Offringa}, {McKinley}, {Hurley-Walker},
  {Briggs}, {Wayth}, {Kaplan}, {Bell}, {Feng}, {Neben}, {Hughes}, {Rhee},
  {Murphy}, {Bhat}, {Bernardi}, {Bowman}, {Cappallo}, {Corey}, {Deshpande},
  {Emrich}, {Ewall-Wice}, {Gaensler}, {Goeke}, {Greenhill}, {Hazelton},
  {Hindson}, {Johnston-Hollitt}, {Jacobs}, {Kasper}, {Kratzenberg}, {Lenc},
  {Lonsdale}, {Lynch}, {McWhirter}, {Mitchell}, {Morales}, {Morgan},
  {Kudryavtseva}, {Oberoi}, {Ord}, {Pindor}, {Procopio}, {Prabu}, {Riding},
  {Roshi}, {Shankar}, {Srivani}, {Subrahmanyan}, {Tingay}, {Waterson},
  {Webster}, {Whitney}, {Williams}, \& {Williams}}]{2014MNRAS.444..606O}
{Offringa}, A.~R., {McKinley}, B., {Hurley-Walker}, N., {et~al.} 2014, \mnras,
  444, 606

\bibitem[{{Offringa} \& {Smirnov}(2017)}]{2017MNRAS.471..301O}
{Offringa}, A.~R. \& {Smirnov}, O. 2017, \mnras, 471, 301

\bibitem[{{Orr{\`u}} {et~al.}(2010){Orr{\`u}}, {Murgia}, {Feretti}, {Govoni},
  {Giovannini}, {Lane}, {Kassim}, \& {Paladino}}]{2010A&A...515A..50O}
{Orr{\`u}}, E., {Murgia}, M., {Feretti}, L., {et~al.} 2010, \aap, 515, A50

\bibitem[{{Pacholczyk}(1970)}]{1970ranp.book.....P}
{Pacholczyk}, A.~G. 1970, {Radio astrophysics. Nonthermal processes in galactic
  and extragalactic sources}

\bibitem[{{Perley} \& {Butler}(2017)}]{2017ApJS..230....7P}
{Perley}, R.~A. \& {Butler}, B.~J. 2017, \apjs, 230, 7

\bibitem[{{Perley} {et~al.}(2011){Perley}, {Chandler}, {Butler}, \&
  {Wrobel}}]{2011ApJ...739L...1P}
{Perley}, R.~A., {Chandler}, C.~J., {Butler}, B.~J., \& {Wrobel}, J.~M. 2011,
  \apjl, 739, L1

\bibitem[{{Perley} \& {Taylor}(1991)}]{1991AJ....101.1623P}
{Perley}, R.~A. \& {Taylor}, G.~B. 1991, \aj, 101, 1623

\bibitem[{{Rajpurohit} {et~al.}(2020){Rajpurohit}, {Hoeft}, {Vazza}, {Rudnick},
  {van Weeren}, {Wittor}, {Drabent}, {Brienza}, {Bonnassieux}, {Locatelli},
  {Kale}, \& {Dumba}}]{Rajpurohit2020a}
{Rajpurohit}, K., {Hoeft}, M., {Vazza}, F., {et~al.} 2020, \aap, 636, A30

\bibitem[{{Rajpurohit} {et~al.}(2021){Rajpurohit}, {Wittor}, {van Weeren},
  {Vazza}, {Hoeft}, {Rudnick}, {Locatelli}, {Eilek}, {Forman}, {Bonafede},
  {Bonnassieux}, {Riseley}, {Brienza}, {Brunetti}, {Br{\"u}ggen}, {Loi},
  {Rajpurohit}, {R{\"o}ttgering}, {Botteon}, {Clarke}, {Drabent},
  {Dom{\'\i}nguez-Fern{\'a}ndez}, {Di Gennaro}, \&
  {Gastaldello}}]{Rajpurohit2021a}
{Rajpurohit}, K., {Wittor}, D., {van Weeren}, R.~J., {et~al.} 2021, \aap, 646,
  A56

\bibitem[{{Rudnick}(2001)}]{2001ASPC..250..372R}
{Rudnick}, L. 2001, in Astronomical Society of the Pacific Conference Series,
  Vol. 250, Particles and Fields in Radio Galaxies Conference, ed. R.~A.
  {Laing} \& K.~M. {Blundell}, 372

\bibitem[{{Scaife} \& {Heald}(2012)}]{2012MNRAS.423L..30S}
{Scaife}, A. M.~M. \& {Heald}, G.~H. 2012, \mnras, 423, L30

\bibitem[{{Shulevski} {et~al.}(2015){Shulevski}, {Morganti}, {Barthel},
  {Harwood}, {Brunetti}, {van Weeren}, {R{\"o}ttgering}, {White}, {Horellou},
  {Kunert-Bajraszewska}, {Jamrozy}, {Chyzy}, {Mahony}, {Miley}, {Brienza},
  {B{\^\i}rzan}, {Rafferty}, {Br{\"u}ggen}, {Wise}, {Conway}, {de Gasperin}, \&
  {Vilchez}}]{2015A&A...583A..89S}
{Shulevski}, A., {Morganti}, R., {Barthel}, P.~D., {et~al.} 2015, \aap, 583,
  A89

\bibitem[{{Smirnov} \& {Tasse}(2015)}]{2015MNRAS.449.2668S}
{Smirnov}, O.~M. \& {Tasse}, C. 2015, \mnras, 449, 2668

\bibitem[{{Tasse} {et~al.}(2018){Tasse}, {Hugo}, {Mirmont}, {Smirnov},
  {Atemkeng}, {Bester}, {Hardcastle}, {Lakhoo}, {Perkins}, \&
  {Shimwell}}]{2018A&A...611A..87T}
{Tasse}, C., {Hugo}, B., {Mirmont}, M., {et~al.} 2018, \aap, 611, A87

\bibitem[{{Thimm} {et~al.}(1994){Thimm}, {Roeser}, {Hippelein}, \&
  {Meisenheimer}}]{1994A&A...285..785T}
{Thimm}, G.~J., {Roeser}, H.~J., {Hippelein}, H., \& {Meisenheimer}, K. 1994,
  \aap, 285, 785

\bibitem[{{van Diepen} {et~al.}(2018){van Diepen}, {Dijkema}, \&
  {Offringa}}]{2018ascl.soft04003V}
{van Diepen}, G., {Dijkema}, T.~J., \& {Offringa}, A. 2018, {DPPP: Default
  Pre-Processing Pipeline}

\bibitem[{{van Haarlem} {et~al.}(2013){van Haarlem}, {Wise}, {Gunst}, {Heald},
  {McKean}, {Hessels}, {de Bruyn}, {Nijboer}, {Swinbank}, {Fallows},
  {Brentjens}, {Nelles}, {Beck}, {Falcke}, {Fender}, {H{\"o}randel},
  {Koopmans}, {Mann}, {Miley}, {R{\"o}ttgering}, {Stappers}, {Wijers},
  {Zaroubi}, {van den Akker}, {Alexov}, {Anderson}, {Anderson}, {van Ardenne},
  {Arts}, {Asgekar}, {Avruch}, {Batejat}, {B{\"a}hren}, {Bell}, {Bell}, {van
  Bemmel}, {Bennema}, {Bentum}, {Bernardi}, {Best}, {B{\^\i}rzan}, {Bonafede},
  {Boonstra}, {Braun}, {Bregman}, {Breitling}, {van de Brink}, {Broderick},
  {Broekema}, {Brouw}, {Br{\"u}ggen}, {Butcher}, {van Cappellen}, {Ciardi},
  {Coenen}, {Conway}, {Coolen}, {Corstanje}, {Damstra}, {Davies}, {Deller},
  {Dettmar}, {van Diepen}, {Dijkstra}, {Donker}, {Doorduin}, {Dromer}, {Drost},
  {van Duin}, {Eisl{\"o}ffel}, {van Enst}, {Ferrari}, {Frieswijk}, {Gankema},
  {Garrett}, {de Gasperin}, {Gerbers}, {de Geus}, {Grie{\ss}meier}, {Grit},
  {Gruppen}, {Hamaker}, {Hassall}, {Hoeft}, {Holties}, {Horneffer}, {van der
  Horst}, {van Houwelingen}, {Huijgen}, {Iacobelli}, {Intema}, {Jackson},
  {Jelic}, {de Jong}, {Juette}, {Kant}, {Karastergiou}, {Koers}, {Kollen},
  {Kondratiev}, {Kooistra}, {Koopman}, {Koster}, {Kuniyoshi}, {Kramer},
  {Kuper}, {Lambropoulos}, {Law}, {van Leeuwen}, {Lemaitre}, {Loose}, {Maat},
  {Macario}, {Markoff}, {Masters}, {McFadden}, {McKay-Bukowski}, {Meijering},
  {Meulman}, {Mevius}, {Middelberg}, {Millenaar}, {Miller-Jones}, {Mohan},
  {Mol}, {Morawietz}, {Morganti}, {Mulcahy}, {Mulder}, {Munk}, {Nieuwenhuis},
  {van Nieuwpoort}, {Noordam}, {Norden}, {Noutsos}, {Offringa}, {Olofsson},
  {Omar}, {Orr{\'u}}, {Overeem}, {Paas}, {Pand ey-Pommier}, {Pandey}, {Pizzo},
  {Polatidis}, {Rafferty}, {Rawlings}, {Reich}, {de Reijer}, {Reitsma},
  {Renting}, {Riemers}, {Rol}, {Romein}, {Roosjen}, {Ruiter}, {Scaife}, {van
  der Schaaf}, {Scheers}, {Schellart}, {Schoenmakers}, {Schoonderbeek},
  {Serylak}, {Shulevski}, {Sluman}, {Smirnov}, {Sobey}, {Spreeuw}, {Steinmetz},
  {Sterks}, {Stiepel}, {Stuurwold}, {Tagger}, {Tang}, {Tasse}, {Thomas},
  {Thoudam}, {Toribio}, {van der Tol}, {Usov}, {van Veelen}, {van der Veen},
  {ter Veen}, {Verbiest}, {Vermeulen}, {Vermaas}, {Vocks}, {Vogt}, {de Vos},
  {van der Wal}, {van Weeren}, {Weggemans}, {Weltevrede}, {White}, {Wijnholds},
  {Wilhelmsson}, {Wucknitz}, {Yatawatta}, {Zarka}, {Zensus}, \& {van
  Zwieten}}]{2013A&A...556A...2V}
{van Haarlem}, M.~P., {Wise}, M.~W., {Gunst}, A.~W., {et~al.} 2013, \aap, 556,
  A2

\bibitem[{{van Weeren} {et~al.}(2012){van Weeren}, {R{\"o}ttgering}, {Intema},
  {Rudnick}, {Br{\"u}ggen}, {Hoeft}, \& {Oonk}}]{vanWeeren2012a}
{van Weeren}, R.~J., {R{\"o}ttgering}, H.~J.~A., {Intema}, H.~T., {et~al.}
  2012, \aap, 546, A124

\bibitem[{{van Weeren} {et~al.}(2020){van Weeren}, {Shimwell}, {Botteon},
  {Brunetti}, {Br{\"u}ggen}, {Boxelaar}, {Cassano}, {Di Gennaro},
  {Andrade-Santos}, {Bonnassieux}, {Bonafede}, {Cuciti}, {Dallacasa}, {de
  Gasperin}, {Gastaldello}, {Hardcastle}, {Hoeft}, {Kraft}, {Mandal},
  {Rossetti}, {R{\"o}ttgering}, {Tasse}, \& {Wilber}}]{Weeren2020}
{van Weeren}, R.~J., {Shimwell}, T.~W., {Botteon}, A., {et~al.} 2020, arXiv
  e-prints, arXiv:2011.02387

\bibitem[{{Wright}(2006)}]{2006PASP..118.1711W}
{Wright}, E.~L. 2006, \pasp, 118, 1711

\end{thebibliography}
